\documentclass[prd,superscriptaddress,amsfonts,amssymb,onecolumn,amsmath,showpacs]{revtex4-2}
\usepackage{bm}
\usepackage{amsfonts}
\usepackage{latexsym}
\usepackage[utf8]{inputenc}
\usepackage{graphicx}
\usepackage{amsmath}
\usepackage{palatino}
\usepackage{mathpazo}
\usepackage[british]{babel}
\usepackage{hhline}
\usepackage{multirow}
\usepackage{textcomp}
\usepackage{float}
\usepackage{booktabs}
\usepackage{dcolumn}
\usepackage{lipsum} 
\usepackage{mdframed}
\usepackage[]{mdframed}
\usepackage{hhline}
\usepackage{multirow}
\usepackage{ragged2e}
\usepackage{hyperref}
\hypersetup{colorlinks,citecolor=blue}
\hypersetup{colorlinks=true,linkcolor=red,filecolor=magenta,    urlcolor=cyan}
\usepackage{amsmath}
\usepackage{xcolor}
\usepackage{orcidlink}
\usepackage{epsfig}
\usepackage{caption}
\usepackage{subcaption}
\usepackage{commath}
\def\jnl@style{\it}
\def\aaref@jnl#1{{\jnl@style#1}}

\def\aaref@jnl#1{{\jnl@style#1}}

\def\aj{\aaref@jnl{AJ}}                   % Astronomical Journal
\def\apj{\aaref@jnl{ApJ}}                 % Astrophysical Journal
\def\apjl{\aaref@jnl{ApJ}}                % Astrophysical Journal, Letters
\def\apjs{\aaref@jnl{ApJS}}               % Astrophysical Journal, Supplement
\def\apss{\aaref@jnl{Ap\&SS}}             % Astrophysics and Space Science
\def\aap{\aaref@jnl{A\&A}}                % Astronomy and Astrophysics
\def\aapr{\aaref@jnl{A\&A~Rev.}}          % Astronomy and Astrophysics Reviews
\def\aaps{\aaref@jnl{A\&AS}}              % Astronomy and Astrophysics, Supplement
\def\mnras{\aaref@jnl{Mon.~Not.~Roy.~Astron.~Soc.}}             % Monthly Notices of the RAS
\def\prd{\aaref@jnl{Phys.~Rev.~D}}        % Physical Review D
\def\prc{\aaref@jnl{Phys.~Rev.~C}}  % Physical Review C
\def\prl{\aaref@jnl{Phys.~Rev.~Lett.}}    % Physical Review Letters
\def\qjras{\aaref@jnl{QJRAS}}             % Quarterly Journal of the RAS
\def\skytel{\aaref@jnl{S\&T}}             % Sky and Telescope
\def\ssr{\aaref@jnl{Space~Sci.~Rev.}}     % Space Science Reviews
\def\zap{\aaref@jnl{ZAp}}                 % Zeitschrift fuer Astrophysik
\def\nat{\aaref@jnl{Nature}}              % Nature
\def\aplett{\aaref@jnl{Astrophys.~Lett.}} % Astrophysics Letters
\def\apspr{\aaref@jnl{Astrophys.~Space~Phys.~Res.}} % Astrophysics Space Physics Research
\def\physrep{\aaref@jnl{Phys.~Rep.}}      % Physics Reports
\def\physscr{\aaref@jnl{Phys.~Scr}}       % Physica Scripta
\def\commat{\aaref@jnl{Comm.~Math.~Phys.}}              % Communications in Mathematical Physics
\def\science{\aaref@jnl{Science}}               % Science
\def\cqg{\aaref@jnl{Classical Quant.~Grav.}}            % Classical and Quantum Gravity
\def\jpcs{\aaref@jnl{JPCS}}                                     % Journal of Physics Conference Series
\def\ijmpd{\aaref@jnl{Int.~J.~Mod.~Phys.~D}}                    % International Journal of Modern Physics D
\def\grg{\aaref@jnl{Gen.~Relat.~Gravit.}}               % General Relativity and Gravitation
\def\rpp{\aaref@jnl{Rep.~Prog.~Phys.}}          % Reports on Progress in Physics
\def\npa{\aaref@jnl{Nucl.~Phys.~A}}        % Nuclear Physics A
\def\lrr{\aaref@jnl{Living Rev.~Rel.}}                   % Living reviews in relativity
\def\jcap{\aaref@jnl{J.~Cosmology Astropart.~Phys.}}    % Journal of cosmology and astroparticle physics
\def\rmp{\aaref@jnl{Rev.~Mod.~Phys.}}   %Reviews of modern physics
\def\epjc{\aaref@jnl{Eur.~Phys.~J.~C}} 
\def\plb{\aaref@jnl{~Phy.~Lett.~B}} 
\def\mpla{\aaref@jnl{Mod.~Phy.~Lett.~A}} 
\def\arxiv{\aaref@jnl{arxiv.org}}

\allowdisplaybreaks[1]
\begin{document}

\title{Dynamical Complexity in Teleparallel Gauss- Bonnet Gravity}

\author{S. A. Kadam\orcidlink{0000-0002-2799-7870}}
\email{k.siddheshwar47@gmail.com}
\affiliation{Department of Mathematics, Birla Institute of Technology and Science-Pilani, Hyderabad Campus, Hyderabad-500078, India}

\author{Santosh V Lohakare\orcidlink{0000-0001-5934-3428}}
\email{lohakaresv@gmail.com}
\affiliation{Department of Mathematics, Birla Institute of Technology and Science-Pilani, Hyderabad Campus, Hyderabad-500078, India}

\author{B. Mishra\orcidlink{0000-0001-5527-3565}}
\email{bivu@hyderabad.bits-pilani.ac.in}
\affiliation{Department of Mathematics, Birla Institute of Technology and Science-Pilani, Hyderabad Campus, Hyderabad-500078, India}

%\date{\today}

\begin{abstract} {\textbf{Abstract:}} The stable critical points and their corresponding cosmology are derived in the teleparallel gravity with an added Gauss-Bonnet topological invariant term. We have analyzed the dynamics of the Universe by presenting two cosmological viable models, showing the potential to describe different phases of the evolution of the Universe. The value of the deceleration parameter ($q$), total equation of state parameter ($\omega_{tot}$) and dark energy equation of state parameter ($\omega_{DE}$) have been presented against each critical point. The existence and stability conditions are also presented. We study the behavior of the phase space trajectories at each critical point. Finally, the evolutionary behavior of the deceleration parameter and the equation of state parameters have been assessed with the initial condition of the dynamical variables, and compatibility has been observed in connection with the present cosmological scenario.    
\end{abstract}

\maketitle


\section{Introduction}
The discovery of an accelerated expansion of the Universe has been confirmed by multiple cosmological observations \cite{Riess:1998cb, Perlmutter:1998np}. Post-discovery, there has been a significant amount of attention to the modified theories of gravity. The source responsible for this cosmic acceleration is an unidentified repulsive form of energy known as dark energy (DE) \cite{Copeland:2006wr, Bamba:2012cp}. To explain the nature of DE along with dark matter (DM) \cite{Baudis:2016qwx, Bertone:2004pz} and inflation \cite{kobayashi:2011}, researchers have attempted modifications to general relativity (GR). We know that it is possible to define a large number of connections in a manifold, and one of such connections is the Levi-Civita connection \cite{misner1973gravitation, nakahara2003geometry}, which is used to construct GR. This Levi-Civita connection is based on the assumptions of torsion-free, metric compatibility, and the assumption of curvature free. The assumption of curvature-free leads to the teleparallel theory of gravity. In the construction of this theory, the Weitzenb\"{o}ck connection \cite{Hayashi:1979qx, Aldrovandi:2013wha} is used, which is different from the Levi-Civita connection. Teleparallel gravity (TG) is an equivalent formulation of GR in which torsion rather than curvature is responsible for the gravitational interaction \cite{Weitzenbock1923, bahamonde:2021teleparallel}.  The TG is a gauge theory, whereas GR is a geometric theory \cite{Clifton:2011jh, bahamonde:2021teleparallel}. The same equations of motion apply to both the theory and the name Teleparallel Equivalent of General Relativity (TEGR) \cite{Aldrovandi:2013wha, Cai:2015emx, bahamonde:2021teleparallel}. In a formulation like TEGR, the gravitational Lagrangian is referred to as the torsion scalar $T$ and is derived from the contraction of the torsion tensor \cite{Krssak:2018ywd, Bahamonde:2015zma, Farrugia:2016qqe}. This is analogous to the Lagrangian of the GR, which is denoted by the curvature scalar $R$ and is built from the contraction of the curvature tensor \cite{Clifton:2011jh}.

The simplest modification to GR is to generalize the action using an arbitrary function of the Ricci scalar $R$, resulting in ghost-free $f(R)$ modified gravity \cite{Sotiriou:2008rp, Faraoni:2008mf, Capozziello:2011et}. In addition to this simple modification of GR, it is possible to construct action with higher curvature corrections, such as the Gauss-Bonnet combination $G$ \cite{WHEELER:198673} and an arbitrary functions $f(G)$ \cite{Nojiri_2005_631, Cognola_2006_73, Felice:2009, jawad:2013}. In TG, the Lagrangian takes the form of the torsion scalar $T$; the equations are equivalent to GR. Teleparallel gravity replaces $T$ with a generic function of torsion as $f(T)$ in TEGR, much as $f(R)$ gravity replaces $R$ with $f(R)$ in Einstein-Hilbert general relativity. The modified TEGR is known as $f(T)$ gravity \cite{Ferraro:2006jd, Ferraro:2008ey, Bengochea:2008gz}. This modification extends the teleparallel Lagrangian $T$ to an arbitrary function $f(T)$. Although TEGR and GR are equivalent at the level of evolution equations, $f(T)$ and $f(R)$ are not equivalent. The cosmological effects of $f(T)$ gravity are therefore novel and intriguing \cite{Linder:2010py, Chen:2010va, Bahamonde:2019zea, Franco:2020lxx}. Inspired by the Gauss-Bonnet modification of GR in \cite{Kofinas:2014owa}, a new generalization of the standard $f(T)$ gravity was proposed. This theory extends the function $f(T)$ to $f(T, T_G)$, where $T_G$ is the Gauss–Bonnet topological invariant. In the teleparallel formalism, i.e. in $f (T, T_G)$ theory, all the degrees of freedom related to torsion are given and then completely extend the $f (T)$ gravity \cite{Capozziello:2016eaz}. 

In this study, we have used the method of dynamical system analysis \cite{Franco:2021, Hohmann:2017, Kofinas:2014daa, dutta:2019, duchaniya2023dynamical}, which is an important tool in cosmology to qualitatively assess the behavior of solutions of the model rather than determining the analytical solution from the complex equations \cite{Hohmann:2017}. This method examines the local asymptotic behavior of critical points of the dynamical system and then connects them to the primary cosmological epochs of the Universe. With this, one can also find a description of the overall dynamics of the Universe \cite{Duchaniya:2022rqu, Kofinas:2014aka, Kadam:2022lgq}. For instance, the radiation and matter-dominated eras correspond to saddle points, whereas a late-time DE domination corresponds to a stable critical point.  However, the dependency on the selection of variables that characterize the solution is linked to the critical, which is one of the limitations of this approach. It is to be noted that the absence of any particular cosmological epoch of the Universe may not indicate the failure of the theory; rather, it is the failure of the associated dynamical system to demonstrate the presence of an epoch \cite{dutta:2019}. Although dynamical systems have been useful in revealing the main features of solutions in specific models, limited attempts have been made for a more systematic analysis of generic $f(T, T_G)$ cosmology \cite{Kofinas:2014aka}. Hence, we aim to derive a general expression for a dynamical system that can explain the de-Sitter, radiation-dominated, as well as cold dark matter-dominated fixed points. To mention here, the dynamical system analysis in curvature-based gravity with the Gauss-Bonnet term, i.e. $f(R, G)$ gravity, has been analysed for the mixed-power law model, highlighting the capability of obtaining eight fixed points that describe radiation-dominated, matter-dominated, and DE-dominated epochs of the evolution of the Universe \cite{Santos2018, Lohakare_2303.14575}.  The development of $n-$ dimensional cosmology for $f(G)$ gravity using the Noether symmetry approach, encompassing power-law models, demonstrating equivalence to GR without imposing $R + f(G)$, presenting solutions like de-Sitter by non-minimal coupling of $f(G)$ to a scalar field, allowing exploration of inflation and DE, concluding with the creation of Hamiltonian formalism and quantum cosmology for the model \cite{Bajardi_2020_80}. An analysis of the topological Gauss-Bonnet scalar in various teleparallel formulations, emphasizing its value for symmetries in GR and the relevance of ``pseudo-invariant" theories in teleparallel Gauss-Bonnet gravity \cite{Bajardi_2308.03632}. We shall analyse this in the torsion-based gravity with the Gauss-Bonnet term, the $f(T, T_G)$ gravity, in the mixed and sum of the separated power-law form.

The exact cosmological solutions in teleparallel Gauss-Bonnet extension theory, with the selection of cosmological viable models, are obtained in Ref. \cite{Capozziello:2016eaz, Kadam2023}. The cosmological bouncing solutions are highlighted in the $f(T.T_G)$ formalism in Ref. \cite{delaCruz-Dombriz:2018nvt}. In the $f(T, T_G)$  gravity formalism, the reconstruction of the cosmological model is studied in \cite{delaCruz-Dombriz:2017lvj}. This theory also shows a confrontation with current observational data \cite{LOHAKARE2023101164}. The dynamical system is analyzed, and the field equations of $f(T, T_G)$  gravity are presented in Ref. \cite{Kofinas:2014aka}. The same approach has been used to analyze the phases of evolution of the Universe in different teleparallel settings such as in $f(T)$ gravity \cite{Duchaniya:2022rqu, Mirza_2017}, $f(T,\phi)$ gravity \cite{Otalora:2013dsa}, $f(T, B)$ gravity \cite{Franco:2020lxx, Franco:2021} and the other modified gravity theories \cite{Narawade:2022a, Agrawal:2023}. In this work, we shall analyze the evolution of the Universe through the dynamical system approach by obtaining the critical points and the behavior of the evolution parameters. We will explain below how Gauss–Bonnet invariants derived from curvature differ from topological invariants derived from torsion in less than a total derivative. Then, the dynamical information in both representations is the same. This result indicates that the topological invariant can also regularize teleparallel torsion. The paper is organized as follows: We first start with an introduction to TG in Section~\ref{sec:intro_tg}, where we review the background properties up to $f(T, T_G)$ gravity. In Section~\ref{Dynamicalanalysis}, two forms of $f(T, T_G)$ have been considered Ref. \cite{Capozziello:2016eaz}. Then, in Subsections \ref{Model-I} and \ref{Model-II}, we introduce the dynamical system analysis for the mixed power law model and the sum of the separated power law model, respectively. We summarize our results in Section~\ref{sec:conclusion} and give our conclusion.

\section{Teleparallel Gravity} \label{sec:intro_tg}

Instead of the Levi-Civita connection, which causes curvature in GR, the teleparallel connection \cite{bahamonde:2021teleparallel} is used in TG.  In GR, the curvature is a product of the Levi-Civita connection $\mathring{\Gamma}^{\sigma}_{\ \ \mu\nu}$ (Over-circles indicate quantities calculated using Levi-Civita connections throughout), which has been replaced by the teleparallel connection $\Gamma^{\sigma}_{\ \ \mu\nu}$ \cite{Hayashi:1979qx, Aldrovandi:2013wha}. In TG, the dynamical variables are the tetrad fields or the vierbein $e^{A}_{\mu}$, where $E_{A}^{\ \ \mu}$ is the inverse of $e^{A}_{\mu}$. The metric tensor $g_{\mu\nu}$ can be calculated from the tetrad fields as follows,
\begin{align}\label{metric_tetrad_rel}
    e^{a}_{\mu} e^{b}_{\nu}\eta_{ab}=g_{\mu\nu}\,,& &  E_{a}^{\ \ \mu} E_{b}^{\ \ \nu}g_{\mu\nu}=\eta_{ab} \,,
\end{align}
The Latin indices denote coordinates on the tangent space, and the Greek indices denote the indices on the general manifold \cite{Cai:2015emx}. Thus, tetrads connect both the spaces and play an important role in raising and lowering indices between the different spaces \cite{Chandrasekhar:1984siy}. TG replaces curvature with torsion, and the torsion tensor can be written exclusively in tetrad fields. This simplifies imposing and verifying vanishing curvature constraints, which has theoretical consequences. Without spinors, space-time formalism may describe TG. So, the curvature-based and TG do not suppress tetrads. They help mathematically describe the gravitational field and its interaction with matter, introduce spinors, and examine curvature or torsion features in both frameworks \cite{Jimenez_2018_jcap}. This could be due to the fact that they are not flat in these settings, whereas in TG, they are introduced as flat connections. Another important feature of these tetrads is that they satisfy the orthogonality requirements,
\begin{align}
    e^{a}_{\ \ \mu} E_{b}^{\ \ \mu}=\delta^a_b\,,&  & e^{a}_{\ \ \mu} E_{a}^{\ \ \nu}=\delta^{\nu}_{\mu}\,.
\end{align}

The teleparallel connection \cite{Weitzenbock1923, Krssak:2015oua} can be defined as,
\begin{equation}
E_{a}^{\ \ \sigma}\left(\partial_{\mu} e^{a}_{\ \ \nu} +  \omega^{a}_{\ \ b\mu} e^{b}_{\ \ \nu}\right) = \Gamma^{\sigma}_{\ \ \nu\mu}  \,,
\end{equation}
The local Lorentz transformation (LLT), wherein $\Lambda^a_b$ represents Lorentz boosts and rotations, is referred to describe the spin connection as $\omega^{a}_{\ \ b\mu} = \Lambda^a_c \partial_{\mu} \Lambda^b_c$ \cite{Hohmann_2018_97_covariant, Golovnev_2017_34, Hohmann_2018_97_teleparallel, Aldrovandi:2013wha}. For a particular metric, there is an infinite number of tetrad choices, which can satisfy Eq. (\ref{metric_tetrad_rel}). So, the spin connection helps counterbalance inertial effects, and in the case of TG, it remains covariant \cite{Krssak:2015oua}. There is a Lorentz frame such that the spin connection is zero. This has been referred to as the Weitzenb\"{o}ck gauge \cite{Weitzenbock1923}. So, from here onward, the spin connection will be dropped following the application of the above-mentioned gauge. Given that the teleparallel Riemann tensor disappears, we define a torsion tensor by \cite{Hayashi:1979qx}
\begin{equation}
    2 \Gamma^{\sigma}_{\ \ [\nu\mu]} =T^{\sigma}_{\ \ \mu\nu}\,,
\end{equation}

The torsion tensor measures the antisymmetry of the connection \cite{Aldrovandi:2013wha}, and the square bracket is the antisymmetry operator. We can also define a contortion tensor as,
\begin{equation}
    \Gamma^{\sigma}_{\mu\nu} - \mathring{\Gamma}^{\sigma}_{\mu\nu} =\frac{1}{2}\left(T_{\mu \ \ \nu}^{\ \ \sigma}+ T_{\nu\ \ \mu}^{\ \ \sigma} - T^{\sigma}_{\ \ \mu\nu}\right) = K^{\sigma}_{\ \ \mu\nu}\,,
\end{equation}
which is directly related to the Levi-Civita connection and plays an important role in defining some scalars and relating curvature and torsional quantities. Using the torsion tensor, a torsion scalar can be defined as \cite{Cai:2015emx, Aldrovandi:2013wha, bahamonde:2021teleparallel}
\begin{equation}\label{eq:torsion_scalar_def}
   \frac{1}{4}T^{\alpha}_{\ \ \mu\nu}T_{\alpha}^{\ \ \mu\nu} + \frac{1}{2}T^{\alpha}_{\ \ \mu\nu}T^{\nu\mu}_{\ \ \  \  \alpha} - T^{\alpha}_{\ \ \mu\alpha}T^{\beta\mu}_{\ \ \ \ \beta} =T\,,
\end{equation}
The action is based on the torsion scalar and will produce the same field equations as the Einstein-Hilbert action. Again, we mention that the Ricci scalar disappears when calculating with the teleparallel connection, so $R \equiv 0$. Here, we can define the standard Ricci scalar, $\mathring{R}=\mathring{R}(\mathring{\Gamma}^{\sigma}_{\ \ \mu\nu})$ \cite{Bahamonde:2015zma, Farrugia:2016qqe} as
\begin{equation}\label{LC_TG_conn}
    \mathring{R} + T - B =R= 0\,.
\end{equation}
$B$ represents a total divergence term and is defined as
\begin{equation}\label{eq:boundary_term_def}
     \frac{2}{e}\partial_{\rho}\left(e T^{\mu \ \ \rho}_{\ \ \mu}\right)=B \,,
\end{equation}
with $e=\det\left(e^{a}_{\ \ \mu}\right)=\sqrt{-g}$ is the determinant of the tetrad. The expression in Eq. (\ref{LC_TG_conn}) conclude that GR and TEGR produce equivalent field equations in the classical regime; hence, both are dynamically equivalent. Another interesting scalar invariant is the Gauss-Bonnet term \cite{Kofinas:2014daa, Kofinas:2014owa, Zubair:2015yma, delaCruz-Dombriz:2017lvj, delaCruz-Dombriz:2018nvt, LOHAKARE2023101164}
\begin{equation}
   \mathring{R}^{2} - 4\mathring{R}_{\mu\nu}\mathring{R}^{\mu\nu} + \mathring{R}_{\mu\nu\alpha\beta}\mathring{R}^{\mu\nu\alpha\beta} =G \,,
\end{equation}
which has been derived in the TG setting and can be defined as
\begin{align}\label{eq:T_G_def}
    T_G &= \Big(K_{a \ \ e}^{\ \ i} K_{b}^{\ \ ej}K_{c \ \ f}^{\ \ k} K_{d}^{\ \ fl} - 2K_{a}^{\ \ ij} K_{b \ \ e}^{\ \ k} K_{c \ \ f}^{\ \ e} K_{d}^{\ \ fl}  + 2K_{a}^{\ \ ij}K_{b \ \ e}^{\ \ k}K_{f}^{\ \ el}K_{d \ \ c}^{\ \ f} + 2K_{a}^{\ \ ij}K_{b \ \ e}^{\ \ k} K_{c,d}^{\ \ \ \ el} \Big)\delta_{ijlk}^{abcd}
\end{align}

where $\delta_{ijkl}^{abcd} = \epsilon^{abcd}\epsilon_{ijkl}$ is the generalized Kronecker delta function \cite{Bahamonde:2016kba}, and comma denotes the differentiation. This has equivalency with the regular Gauss-Bonnet term up to a total divergence term defined as follows,
\begin{equation}
    \frac{1}{e}\delta^{abcd}_{ijkl}\partial_{a} \left[K_{b}^{\ \ ij}\left(K_{c\ \ ,d}^{\ \ kl} + K_{d \ \ c}^{\ \ m}K_{m}^{\ \ kl}\right)\right]=B_G  \,.
\end{equation}
The $T_G, B_G$ together produce the teleparallel equivalent of the Gauss-Bonnet term as
\begin{equation}
     -T_G + B_G=G \,.
\end{equation}

Similarly to the modifications of GR, we consider a generalization of the Einstein-Hilbert action by adding generalizations in terms of both the torsion scalar and the Gauss-Bonnet term. The Gauss-Bonnet term has been linked to both inflationary and dynamical aspects of DE. We consider the action equation as \cite{Kofinas:2014daa, Kofinas:2014owa}
\begin{equation}\label{f_T_G_ext_Lagran}
    \mathcal{S}_{F(T,T_G)}^{} =  \frac{1}{2\kappa^2}\int \mathrm{d}^4 x\; e\,F(T,T_G) + \int \mathrm{d}^4 x\; e\mathcal{L}_{\text{m}}\,,
\end{equation}

where $\kappa^2=8\pi G$, and $\mathcal{L}_{\text{m}}$ is the matter Lagrangian in the Jordan frame. We wish to mention that the $f(T)$ theory has only first-order derivatives of the tetrad in action, whereas $f(T, T_G)$ action comes with second-order derivatives of the tetrad. However, it is beyond the scope of the present investigation to consider the possibility of Ostrogradsky instability. In this formalism, the second and fourth-order contributions to the field equations are associated with the torsion scalar and Gauss-Bonnet term contributions, respectively. For a spatially flat, homogeneous, and isotropic cosmological background, the tetrad can be written as \cite{bahamonde:2021teleparallel}
\begin{equation}\label{flrw_tetrad}
    e^{A}_{\ \ \mu}=\textrm{diag}(1,a(t),a(t),a(t))\,,
\end{equation}
where $a(t)$ is the scale factor that produces the spatially flat homogeneous and isotropic metric
\begin{equation}
    ds^2 = -dt^2+a(t)^2(dx^2+dy^2+dz^2)\,,
\end{equation}
 Using the definitions in Eqns. \eqref{eq:torsion_scalar_def}-- \eqref{eq:T_G_def}, the torsion scalar and the teleparallel equivalent of the Gauss-Bonnet term can be calculated as,
\begin{equation}\label{T,T_G}
    T =6H^2\,,\quad T_G = 24H^2\left(\dot{H}+H^2\right),\,
\end{equation}
where the Gauss-Bonnet term turns out to have the same value for this background as its curvature analogue. The Friedmann equations for this set-up obtained to be \cite{Kofinas:2014daa, Bahamonde:2016kba}

\begin{subequations}
\begin{align}
   F - 12 H^2 F_T - T_G F_{T_G} + 24 H^3 \dot{F}_{T_G} = 2 \kappa^2 \rho, \label{1stFE}\\ 
   F - 4 (\dot{H} +3 H^2) F_{T} - 4 H \dot{F}_{T}-T_{G} F_{T_{G}}
   + \frac{2}{3 H} T_G \dot{F}_{T_G} + 8 H^2 \ddot{F}_{T_{G}} = -2\kappa^2 p\label{2ndFE}.
\end{align}
\end{subequations}

An over-dot refers to derivatives with respect to cosmic time $t$. To establish the deviation of this theory from GR and to better probe the role from the modified Lagrangian, we consider $F(T, T_G)=-T+f(T, T_G)$. Subsequently, Eqns. (\ref{1stFE})--(\ref{2ndFE}) reduce respectively as,

\begin{subequations}
\begin{align}
   6H^2+f-12H^2 f_{T}-T_{G} f_{T_G}+24H^3 \dot{f}_{T_G}=2\kappa^2 \rho \,,\label{1stfe} \\ 
     2(2\dot{H}+3H^2)+f-4(\dot{H}+3H^2)f_{T}-4H\dot{f}_{T} - T_{G} f_{T_G}+\frac{2}{3H} T_G \dot{f}_{T_G}+8H^2 \ddot{f}_{T_G}=-2\kappa^2 p\,.\label{2ndfe}
\end{align}
\end{subequations}

 These equations will reduce to GR when $f=0$ or $f=T_G$, and in the latter case, as expected, the $T_G$ terms will get cancelled. Now, the Friedmann equations (\ref{1stfe}- \ref{2ndfe}) become

\begin{subequations}
\begin{align}
3H^2=\kappa^2\left(\rho+\rho_{DE}\right)\,,\label{1stfe1}\\
    3H^2+2\dot{H}=-\kappa^2\left(p+p_{DE}\right)\,,\label{2ndfe2}
\end{align}
\end{subequations}
from where the expressions for energy density and pressure for the DE sector can be retrieved as

\begin{subequations}
\begin{align}
    \frac{-1}{2\kappa^2}\left(f-12H^2 f_{T}-T_G f_{T_G}+24H^3 \dot{f}_{T_G}\right)&=\rho_{DE}\,,\label{fede1}\\
    \frac{1}{2\kappa^2}\left(f-4(\dot{H}+3H^2) f_{T}-4H\dot{f}_T-T_G f_{T_G}+\frac{2}{3H} T_G \dot{f}_{T_{G}}+8H^2 \ddot{f}_{T_G}\right)&=p_{DE}\,.\label{fede2}
\end{align}
\end{subequations}
The expression for the EoS parameter for dark energy can be written as, 
\begin{align}
    \omega_{DE}=-1+\frac{8H^2 \ddot{f}_{T_G}-4\dot{H}f_{T}-4H\dot{f}_T + \frac{2}{3H} T_G \dot{f}_{T_G}  -24 H^3 \dot{f}_{T_G}}{12H^2 f_T +T_G f_{T_G} -24 H^3 \dot{f}_{T_G}-f}\,. \label{omegade}
\end{align}
These DE densities and pressure expressions will satisfy the continuity equation.
\begin{align}
    \dot{\rho}_{DE}+3H\left(\rho_{DE}+p_{DE}\right)=0
\end{align}
%%%%%%%%%%
\section{Dynamical System Analysis in \texorpdfstring{$f(T,T_G)$}{} 
Gravity}\label{Dynamicalanalysis}

We shall first define the appropriate dynamical variables to analyze the dynamical system of higher-order teleparallel gravity \cite{Franco:2020lxx, Franco:2021}. By differentiating the dynamical variables with respect to $N=ln(a)$, the expressions for the autonomous dynamical system can be obtained and, subsequently, the critical points. We consider the Universe to be filled with two fluids, $\rho=\rho_{m}+\rho_{r}$, where $\rho_m$ and $\rho_{r}$ respectively denote the energy density for matter and radiation phase. In the matter phase, the matter pressure $p_m=0$ and therefore $\omega_m$ also vanishes. In the radiation phase,  the EoS parameter, $\omega_{r}=\frac{1}{3}$. Therefore, with these considerations, we define the dynamical variables as follows
\begin{align}
X=f_{T_G} H^2 \,,\quad Y=\dot{f}_{T_G}H \,,\quad Z=\frac{\dot{H}}{H^2} \,,\quad 
V=\frac{\kappa^2 \rho_r}{3H^2}\,,\quad
W=-\frac{f}{6H^2}\,.\label{dynamical variables}
\end{align}

The standard density parameters expressions for matter $(\Omega_m)$, radiation $(\Omega_r)$ and DE $(\Omega_{DE})$ phase are respectively,
\begin{align}
\Omega_{m}=\frac{\kappa^2 \rho_{m}}{3H^2}, \quad \Omega_{r}=\frac{\kappa^2 \rho_{r}}{3H^2}, \quad \Omega_{DE}=\frac{\kappa^2 \rho_{DE}}{3H^2},
\end{align}
with 
\begin{align}
\Omega_{m}+\Omega_{r}+\Omega_{DE}=1.
\end{align}
In terms of dynamical variables, we have,
\begin{align}
\Omega_{m}+\Omega_{r}+W+2f_{T}+4XZ+4X-4Y=1,
\end{align}
and
\begin{align}
    \Omega_{DE}=W+2f_{T}+4XZ+4X-4Y.
\end{align}

To express the autonomous dynamical system, we define the parameter $\lambda=\frac{\ddot{H}}{H^3}$ \cite{Odintsov:2018}, and so the general form of the dynamical system can be obtained by differentiating the dimensionless variables with respect to $N=ln(a)$,
 
\begin{align}
\frac{dX}{dN}&=2XZ+Y\,,\nonumber\\
\frac{dY}{dN}&=-\frac{3}{4}-\left(Z+2\right)Y-\frac{Z}{2}+\frac{3W}{4}+\left(Z+3\right)\frac{f_{T}}{2}+\frac{\dot{f}_{T}}{2H}\nonumber +3X\left(Z+1\right)-\frac{V}{4}\,,\nonumber\\
\frac{dZ}{dN}&=\lambda-2Z^2\,,\nonumber\\
\frac{dW}{dN}&=-2f_{T}Z-8XZ^2-4X\lambda-16XZ-2WZ\,,\nonumber\\
\frac{dV}{dN}&=-4V-2VZ\,.\label{generaldynamicalsystem}
\end{align}

To form the autonomous dynamical system, we need a form for $f(T, T_G)$ in which the terms $f_{T},\frac{\dot{f}_{T}}{H}$ can be written in terms of the dimensionless variables, which we shall discuss by considering two forms of $f(T, T_G)$ as two models.
%%%%%%%%%%%%%%%%%%%%%%%%%%%%%%%%%%%

\subsection{Mixed Power Law Model}\label{Model-I}
We consider the mixed power law form of $f(T, T_G)$ \cite{Capozziello:2016eaz} as
\begin{align}
f(T,T_G)=f_{0} T_{G}^{k} T^{m},\label{firstmodel}
\end{align}
where $f_{0}, m, k$ are the arbitrary constants, the GR limit can be recovered for vanishing index powers \cite{Escamilla-Rivera:2019ulu}. The motivation to consider this form is to study the role of parameter $\lambda$ in its most general form. From (\ref{dynamical variables})  and (\ref{firstmodel}), we can write $f_{T}=-m W$, and this will guarantee the autonomous dynamical system and the dynamical variable $X$ becomes, \begin{align}
X&=f_{T_G} H^2=f_{0} k G^{k-1} T^{m} H^{2}\nonumber\\
&=\frac{kfH^2}{G}=\frac{kf}{24(\dot{H}+H^2)}=\frac{kf}{6H^2}\left(\frac{1}{4\left(\frac{\dot{H}}{H^2}+1\right)}\right)=\frac{-Wk}{4(Z+1)}.
\end{align}

The parameter $X$ depends on $W$ with the condition on $Z\ne -1$. This condition also guarantees $T_G=24H^4(Z+1)$, i.e. the non-vanishing teleparallel Gauss-Bonnet term. The dynamical variable $W$ can be rewritten as
\begin{align}
W=-\frac{4X(Z+1)}{k} \label{eqforz}
\end{align}
from Eq. (\ref{eqforz}), an expression for $\lambda$ can be obtained,
\begin{align}
\lambda&=\frac{1}{1-k}\left[2(Z+1) Z (m-2)+2Z^2(k+1)+4kZ- \frac{(Z+1)Y}{X}\right]\label{eqforlambda}
\end{align}

From Eqns. (\ref{eqforz}-\ref{eqforlambda}), one can note that the system equations become singular at $k=0$ or $k=1$, i.e. we can not use it in the limit to $f(T)$ gravity ($m$ arbitrary) and in the limit to GR $(m=0, k=0)$. This is a shortcoming of the chosen variables since the original equations (\ref{1stfe} - \ref{2ndfe}) are regular in these limits. Although the parameter $f_0$ does not play any role in the further analysis, the system equations will converge to GR spontaneously for the case $k=0, m=0$, but since due to the singularity of system equations at $k=0$, we will drop this value in our analysis. We will consider $X$, $Y$, $Z$, $V$ as independent variables, $W$ and $\lambda$ as dependent variables and then inserting $f_{T}$ and $\frac{\dot{f}_{T}}{H}$ in to the system (\ref{generaldynamicalsystem}), the system equation we obtain,

\begin{align}
\frac{dX}{dN}&=2XZ+Y\,,\nonumber\\
\frac{dY}{dN}&=-\frac{k^2 (V-12 X (Z+1)+4 Y (Z+2)+2 Z+3)}{4 (k-1) k}+\frac{-8 X (Z+1) (m (Z-3)+3)+V+2 Z+3}{4 (k-1)}\nonumber\\& \,\,\,\,\,\,\,\, -\frac{4 (2 m-1) X (Z+1) (2 m Z+3)}{4 (k-1) k}+\frac{4 Y (2 m (Z+1)+Z+2)}{4 (k-1)},\nonumber\\
\frac{dZ}{dN}&=\frac{(Z+1) (Y-2 X Z (2 k+m-2))}{(k-1) X},\nonumber\\
\frac{dV}{dN}&=-4V-2VZ\,.\label{DynamicalSystem1stmodel}
\end{align}

We get the density parameter for DE and matter, respectively, in dynamical variables as,
\begin{subequations}
\begin{align}
\Omega_{DE}&=-4 Y+\frac{4 X (Z+1) (k+2 m-1)}{k},  \\
\Omega_{m}&=1 - V + 4 Y -[\frac{4}{k} (-1 + k + 2 m) X (1 + Z)]\,.
\end{align}
\end{subequations}

The critical points of the dynamical system can be obtained by considering $\frac{dX}{dN}=0, \frac{dY}{dN}=0, \frac{dZ}{dN}=0, \frac{dV}{dN}=0$, and are obtained as in Table \ref{TABLE-I}. 

\begin{table*}[!htb]
 % title of Table
\centering % used for centering table
\begin{tabular}{|*{6}{c|}}\hline 
    \parbox[c][0.8cm]{3.3cm}{\textbf{Name of Critical Point}} & $\textbf{X}$ & $\textbf{Y}$ & $\textbf{Z}$ & $\textbf{V}$ & \textbf{Exist for}\\ [0.5ex]\hline \hline % inserts table %heading \hline\hline % inserts a single horizontal line
    \parbox[c][1.2cm]{3cm}{$A_1= (x_1,y_1,z_1,v_1)$} & $x_1$ & $4 x_1$ & $-2$ & $v_1$ & \begin{tabular}{@{}c@{}}$v_{1}-4 x_{1}-1\neq 0$, $m=\frac{v_{1}-12 x_{1}-1}{v_{1}-4 x_{1}-1},$ $k=\frac{1-m}{2},$ \\ $-v_{1} x_{1}+8 x_{1}^2+x_{1}\neq 0$ \end{tabular}\\
    \hline
    \parbox[c][1cm]{3cm}{$A_2= (x_2,y_2,z_2,v_2)$} & $x_{2}$ & $3 x_{2}$ & $-\frac{3}{2 }$ & $0$ &  \begin{tabular}{@{}c@{}}$8 x_{2}+1\neq 0, x_{2}\ne 0,$\\ $k=\frac{1-m}{2},$ $m^2-1\neq 0 $
    \end{tabular}\\
    \hline
    \parbox[c][0.8cm]{3cm}{$A_3= (x_3,y_3,z_3,v_3)$} & $\frac{1}{4}$ & $0$ & $0$ & $0$ & $k^2-k\neq 0,$ $m=\frac{1}{2}$ \\
    \hline
    \parbox[c][1.2cm]{3cm}{$A_4= (x_4,y_4,z_4,v_4)$} & $x_4$ & $0$ & $0$ & $0$ &  \begin{tabular}{@{}c@{}}$4 x_4-1\neq 0,$ $\left(2 m-1\right) x_4 \left(8 x_4+1\right) \left(8 m x_4-1\right)\neq 0,$\\ $k=-\frac{4 (2 m x_4-x_4)}{4 x_4-1}$.\end{tabular}\\
    \hline
    \parbox[c][1.5cm]{3cm}{$A_5= (x_5,y_5,z_5,v_5)$} & $x_5$ & $y_5$ & $-\frac{y_5}{2 x_5}$ & $0$ & \begin{tabular}{@{}c@{}}$y_{5} (12 x_{5}-2 y_{5}+1) (3 x_{5}-y_{5})\neq 0$\\$m=\frac{4 x_{5}+2 y_{5}+1}{12 x_{5}-2 y_{5}+1}$ \\$k=\frac{1-m}{2},$ $m^2-1\neq 0$ \end{tabular}\\
    \hline
\end{tabular}
\caption{The critical points (Model-I)}
\label{TABLE-I}
\end{table*}

The stability properties of the critical points are put into groups: (i) stable node: all the eigenvalues are negative; (ii) unstable node: all the eigenvalues are positive; (iii) saddle-node: one, two, or three of the four eigenvalues are positive, and the remaining are negative; (iv) stable spiral node: The determinant of the Jacobian matrix is negative, and the real part of all the eigenvalues is negative. We have summarized the stability of all the critical points for Model--I in Table \ref{TABLE-II}. To identify the phase of evolution, the value of the deceleration parameter $q$ and the EoS parameters ($\omega_{tot}$, $\omega_{DE}$) are presented, corresponding to each critical point.
\begin{table*}[!htb]
    \centering % used for centering table
    \begin{tabular}{|*{5}{c|}}\hline
    \parbox[c][0.8cm]{1cm}{C. P.} & Stability Conditions & $q$ & $\omega_{tot}$ & $\omega_{DE}$ \\ [0.5ex] % inserts table %heading
    \hline\hline % inserts single horizontal line
    \parbox[c][0.7cm]{1cm}{$A_1$} & Unstable & $1$ & $\frac{1}{3}$ & $\frac{1}{3}$ \\
    \hline
    \parbox[c][1cm]{1cm}{$A_2$} & $\begin{tabular}{@{}c@{}} Stable for\\$m<-1\land \frac{1-m}{6 m-10}<x_2\leq \frac{8 m-8}{9 m^2-48 m+71}$\\ Otherwise Unstable\end{tabular}$ & $\frac{1}{2}$ & $0$ & $0$ \\
    \hline
    \parbox[c][1cm]{1cm}{$A_3$} & $\begin{tabular}{@{}c@{}} Stable for\\$\frac{1}{4}<k\leq \frac{13}{25}$\end{tabular}$  & $-1$ & $-1$ & $-1$\\
    \hline
    \parbox[c][1cm]{1cm}{$A_4$} & $\begin{tabular}{@{}c@{}} Stable for\\$x_4<-\frac{1}{6}\land \left(m\leq \frac{17 x_4+2}{72 x_4^2+24 x_4+2}\lor m>\frac{4 x_4+1}{12x_4+1}\right)$\end{tabular}$  & $-1$ & $-1$ & $-1$\\
    \hline
    \parbox[c][1cm]{1cm}{$A_5$} & $\begin{tabular}{@{}c@{}} Stable for\\$\left(x_5<0\land y_5>3 x_5\right)\lor \left(x_5>0\land y_5<3 x_5\right)$\end{tabular}$  & $-1+\frac{y_5}{2 x_5}$ & $-1+\frac{y_5}{3 x_5}$ & $-1+\frac{y_5}{3 x_5}$\\
    [1ex] % [1ex] adds vertical space
    \hline %inserts a single line
    \end{tabular}
    \caption{Stability condition, deceleration and EoS parameter (Model--I)}
    \label{TABLE-II}
\end{table*}

The cosmological solutions for the corresponding evolution equation at each critical point, along with the standard density parameters ($\Omega_{m}$, $\Omega_{DE}$, $\Omega_{r}$) for Model--I are presented in Table \ref{TABLE-III}.
\begin{table*}[!htb]
     % title of Table
    \centering % used for centering table
    \begin{tabular}{|*{6}{c|}}\hline
    \parbox[c][0.8cm]{1cm}{C. P.} & Evolution Eqs. & Universe phase & $\Omega_{m}$& $\Omega_{r}$ & $\Omega_{DE}$\\ [0.5ex] % inserts table %heading
    \hline\hline % inserts single horizontal line
    \parbox[c][0.7cm]{1cm}{$A_1$} & $\dot{H}=-2 H^{2}$ & $a(t)= t_{0} (2 t+c_{2})^\frac{1}{2}$ & $0$& $v_1$ & $1-v_1$\\
    \hline
    \parbox[c][0.7cm]{1cm}{$A_2$} & $\dot{H}=-\frac{3}{2}H^{2}$ & $a(t)= t_{0} (\frac{3}{2}t+c_{2})^\frac{2}{3}$ & $\frac{2 \left(3 m-5\right) x_2}{m-1}+1$& $0$ & $\frac{2 \left(5-3 m\right) x_2}{m-1}$\\
    \hline
    \parbox[c][0.7cm]{1cm}{$A_3$} & $\dot{H}=0$ & $a(t)= t_{0} e ^{c_{1}t} $ & $0$& $0$ & $1$\\
    \hline
    \parbox[c][0.7cm]{1cm}{$A_4$} &$\dot{H}=0$ & $a(t)= t_{0} e ^{c_{1}t} $ &  $0$& $0$ & $1$\\
    \hline
    \parbox[c][0.7cm]{1cm}{$A_5$} &  $\dot{H}=-\frac{y_5}{2 x_5}H^2$ & $a(t)= t_{0} (\frac{y_5}{2 x_5}t+c_{2})^\frac{2x_5}{y_5}$ & $0$& $0$ & $1$\\
    [1ex] % [1ex] adds vertical space
    \hline %inserts single line
    \end{tabular}
    \caption{Phase of the Universe, density parameters (Model--I)}
     % is used to refer this table in the text
    \label{TABLE-III}
\end{table*}

The eigenvalues for the Jacobian matrix of the dynamical system in Eq. (\ref{DynamicalSystem1stmodel}) at each critical point are presented in Table \ref{TABLE-IV}.

\begin{table*} [!htb]
\begin{tabular}{|*{2}{c|}}\hline \hline
\parbox[c][0.5cm]{0.5cm}{C.P.} & \parbox[c][1cm]{15cm}{Eigenvalues}\\ \hline
\hline
\parbox[c][0.5cm]{0.5cm}{$A_1$} & \parbox[c][1cm]{15cm}{$\left\{0,1,\frac{x_1 \left(v_1-8 x_1-1\right)-r}{2 x_1 \left(-v_1+8 x_1+1\right)},\frac{r+x_1 \left(v_1-8 x_1-1\right)}{2 x_1 \left(-v_1+8 x_1+1\right)}\right\}$}\\ \hline
\parbox[c][0.5cm]{0.5cm}{$A_2$} & \parbox[c][1.5cm]{15cm}{$\left\{0,-1,-\frac{\sqrt{\left(m^2-1\right) x_2 \left(m \left(\left(9 m-48\right) x_2-8\right)+71 x_2+8\right)}}{4 \left(m^2-1\right) x_2}-\frac{3}{4},\frac{\sqrt{\left(m^2-1\right) x_2 \left(m \left(\left(9 m-48\right) x_2-8\right)+71 x_2+8\right)}}{4 \left(m^2-1\right) x_2}-\frac{3}{4}\right\}$}\\ \hline
\parbox[c][0.5cm]{0.5cm}{$A_3$} & \parbox[c][1cm]{15cm}{$\left\{-4,-3,\frac{-\sqrt{25 k^2-38 k+13}-3 k+3}{2 \left(k-1\right)},\frac{\sqrt{25 k^2-38 k+13}-3 k+3}{2 \left(k-1\right)}\right\}$}\\ \hline
\parbox[c][0.5cm]{0.5cm}{$A_4$} & \parbox[c][1cm]{15cm}{$\left\{-4,-3,-\frac{s}{2 x_4 \left(8 m x_4-1\right)}-\frac{3}{2},\frac{3 x_4 \left(1-8 m x_4\right)+s}{2 x_4 \left(8 m x_4-1\right)}\right\}$}\\ \hline
\parbox[c][0.5cm]{0.5cm}{$A_5$} & \parbox[c][1cm]{15cm}{$\left\{0,-\frac{4 x_5-y_5}{x_5},-\frac{-8 x_5 y_5+12 x_5^2+y_5^2}{2 x_5 \left(2 x_5-y_5\right)},-\frac{-5 x_5 y_5+6 x_5^2+y_5^2}{x_5 \left(2 x_5-y_5\right)}\right\}$}\\ 
\hline
\end{tabular}
\caption{Eigenvalues corresponding to each critical point (Model--I)}
\label{TABLE-IV}
\end{table*}
where, 
\begin{align}
r&=\sqrt{-\left(x_1 \left(v_1-8 x_1-1\right) \left(-v_1 \left(9 x_1+2\right)+2 v_1^2+8 x_1^2+x_1\right)\right)}\,,\quad
s&=\sqrt{x_4 \left(8 m x_4-1\right) \left(2 m \left(6 x_4+1\right){}^2-17 x_4-2\right)}\,.\nonumber
\end{align}

\begin{figure}[H]
    \centering
    \includegraphics[width=59mm]{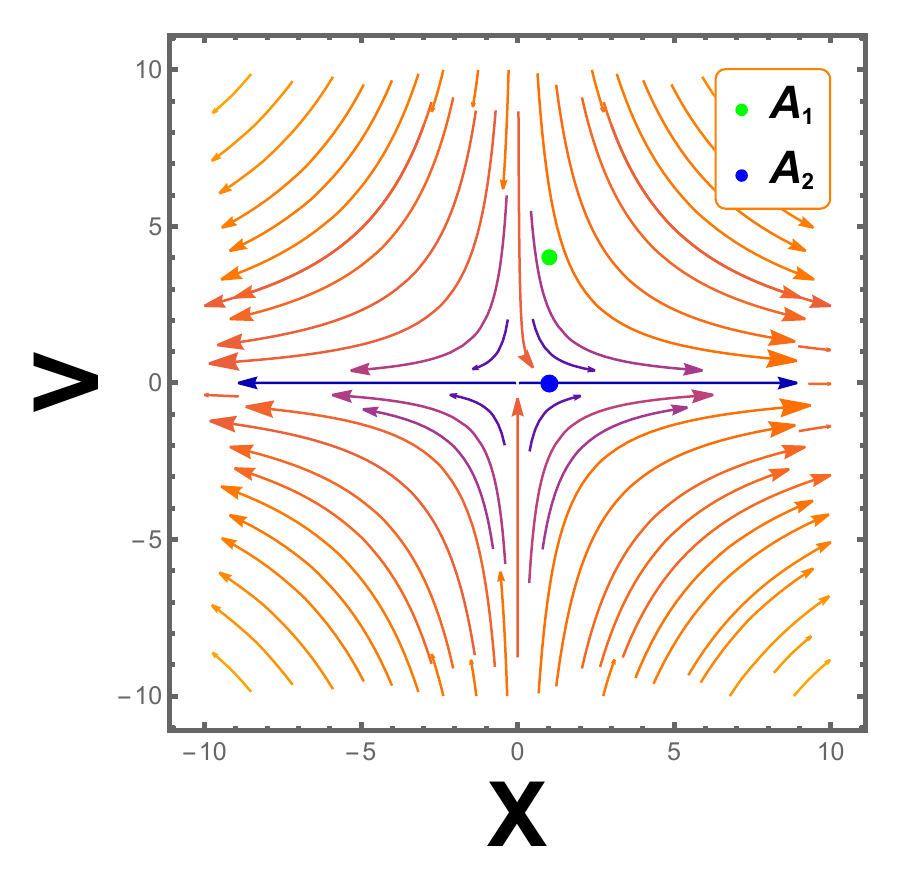}
    \includegraphics[width=59mm]{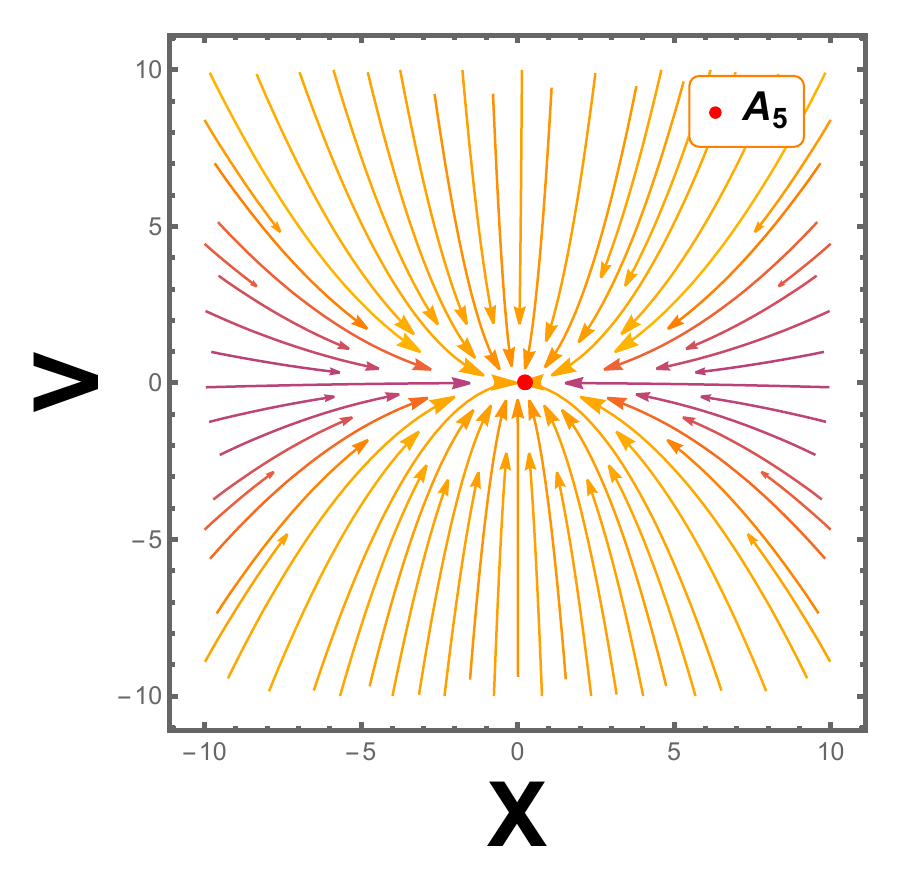}
    \includegraphics[width=59mm]{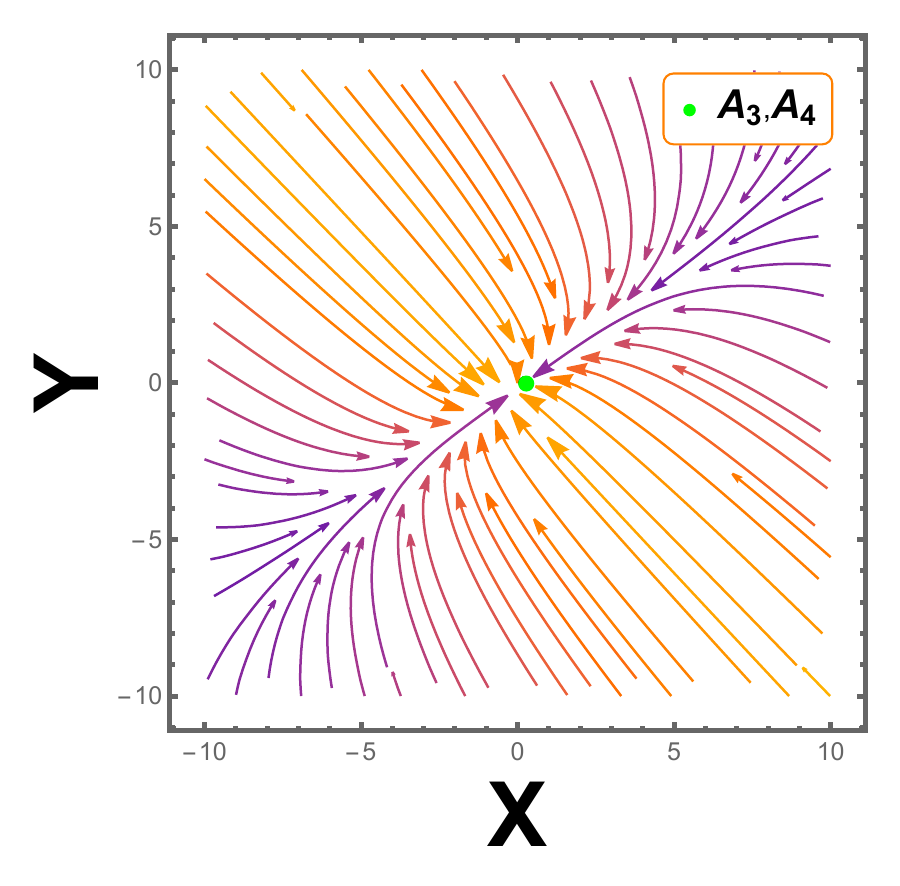}
    \caption{ $2D$ phase portrait with $k=0.029$, $m=0.5$ (Model--I). } \label{Fig1} 
\end{figure}

\textbf{Critical Point $A_1(Z=-2)$:} The critical point $A_{1}$ represents radiation dominated era with $\omega_{tot}=\omega_{DE}=\frac{1}{3}$ and  $q=1$, it describes the standard radiation dominated era for $v_{1}=1$, where $\Omega_{r}=1$ and $\Omega_{m}=0,\Omega_{DE}=0$. From Fig. \ref{Fig1}, it can be analysed that critical point $A_1$ representing radiation-dominated era exists at a parametric range $m=\frac{v_{1}-12 x_{1}-1}{v_{1}-4 x_{1}-1},$ $k=\frac{1-m}{2}$, and is unstable (saddle) due to presence of a positive eigenvalue. From the 2-D phase space in Fig. \ref{Fig1}, we can observe that phase space trajectories move away from the critical point, further confirming the saddle point behavior.

\textbf{Critical Point $A_2(Z=-\frac{3}{2})$:} The critical point $A_{2}$ describes a non-standard cold dark matter-dominated era with the negligible contribution of DE density $\Omega_{DE}=\frac{2 \left(5-3 m\right) x_2}{m-1}$. This critical point will represent a standard cold dark matter-dominated era for $x_2=0$ or $m=\frac{3}{5}$. The CDM-dominated era can also be described at the critical point $A_2$ in the parametric range $k=\frac{1-m}{2},$ $m^2-1\neq 0$. The eigenvalues for the Jacobian matrix at this critical point are non-hyperbolic in nature, as presented in Table \ref{TABLE-IV} and show stability at the condition described in Table \ref{TABLE-II}. This critical point will either be a saddle or unstable node for the parameter range lying outside the stability condition described in Table \ref{TABLE-II}. We have plotted 2-D and 3-D phase portraits for $m=0.5$, for which stability condition on $x_2$ is $-0.0714286<x_2\le -0.0812183$. The coordinate of the phase portraits for 2-D and 3-D is in the unstable range for a critical point $A_2$, hence describing the saddle point nature of this critical point. The power law solution corresponding to the evolution equation at this critical point and the standard density parameters corresponding to different phases of the Universe evolution are presented in Table \ref{TABLE-III}. In this case, the values of $\omega_{tot}=\omega_{DE}=0$, and the deceleration parameter will take the value $q=\frac{1}{2}$ which is positive; hence this critical point can not describe the current cosmic acceleration.  

\textbf{Critical Points $A_3, A_4(Z=0)$:} The critical points $A_3$ and $A_4$ both are the de-Sitter solutions with $\Omega_{DE}=1, \Omega_{m}=0, \Omega_{r}=0$. The value of $\omega_{tot}=\omega_{DE}=q=-1$; hence these two critical points describe the current accelerated expansion of the Universe. The de-Sitter solution can be explained at the critical point $A_3$, and it exists in the parametric range $k^2-k\neq 0,$ $m=\frac{1}{2}$. The critical point $A_4$ is valid in the parametric range $\left(2 m-1\right) x_4 \left(8 x_4+1\right) \left(8 m x_4-1\right)\neq 0,$ $k=-\frac{4 (2 m x_4-x_4)}{4 x_4-1}$. The eigenvalues of the Jacobian matrix at both the critical points are presented in Table \ref{TABLE-IV} and are hyperbolic in nature. The stability conditions are described in Table \ref{TABLE-II}. The phase space trajectories show attractor behavior, which can be observed from Fig. \ref{Fig1}.

\textbf{Critical Point $A_5(\text{$Z$ depends on $Y$, $X$})$:} The value of the deceleration parameter, EoS parameters for this critical point is dependent on coordinates $x$ and $y$ as the value of $q=-1+\frac{y_5}{2 x_5},\omega_{tot}=\omega_{DE}=-1+\frac{y_5}{3 x_5}$. This critical point represents a  DE-dominated era with $\Omega_{DE}=1.$ This can explain the current cosmic acceleration at $\left(x_5<0\land y_5>2 x_5\right)\lor \left(x_5>0\land y_5<2 x_5\right)$. From the eigenvalues presented in Table \ref{TABLE-IV}, we can infer that the critical point is non-hyperbolic in nature and is stable at the stability condition presented in Table \ref{TABLE-II}.  Since the phase space trajectories behavior can be analyzed from Fig. \ref{Fig1} are attracting towards the critical point $A_{5}$, this critical point is an attractor.

\begin{figure}[H]
    \centering
    \includegraphics[width=70mm]{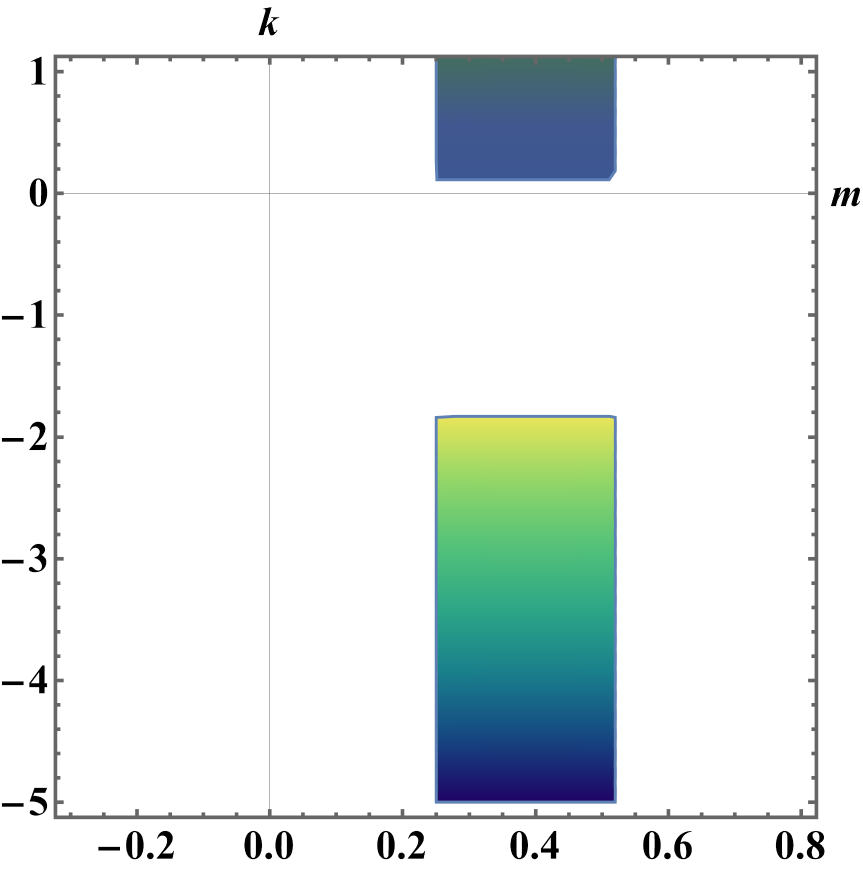}
    \caption{Region Plot for critical points $A_3$ and $A_4$ at $x_4=-\frac{1}{3}$ } \label{regionplotm1}
\end{figure}

In Fig. \ref{regionplotm1}, we have given the region plot, which will be helpful to visualize the ranges of the model parameters $m$ and $k$ at which both the critical points ($A_3, A_4$) are stable and confirms the accelerating behavior. Mathematically, one can obtain the range as, $\left((\frac{1}{4} < k \leq \frac{13}{25}) \wedge \left(m > \frac{1}{6}\right)\right) \vee \left((\frac{1}{4} < k \leq \frac{13}{25}) \wedge \left(m \leq -\frac{11}{6}\right)\right)$.\\

\begin{figure}[H]
    \centering
    \includegraphics[width=70mm]{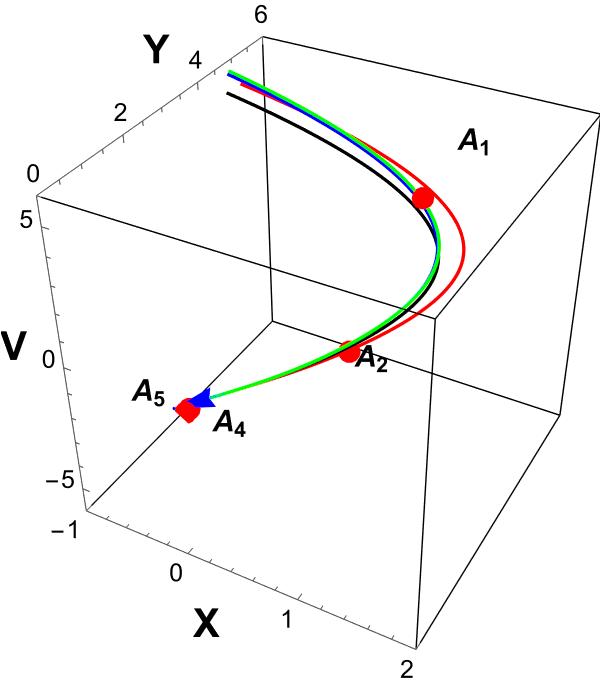}
    \caption{3-D phase portrait with $k=0.029, m=0.5$ (Model--I) } \label{3-D Model-I}
\end{figure}
The 3-D phase portrait presented in Fig. \ref{3-D Model-I} allows us to analyze the behavior of trajectories at the critical points representing different phases of Universe evolution. Phase space trajectory passes through the critical point $A_1 \rightarrow A_2 \rightarrow A_4, A_5.$  We can see from the figure that the chosen trajectory evolves from a radiation-dominant solution corresponding to critical point $A_1$ to an accelerating solution corresponding to critical points $A_4$ and $A_5$. There may be first red dots in the trajectory transition between $A_1$ (radiation) and $A_2$ (matter) and last dots in the trajectory transition between $A_4$ and $A_5$ (de-sitter). Our final attractors' points, $A_4$ and $A_5$ represent the de-sitter epoch with cosmic acceleration.
\begin{figure}[H]
    \centering
    \includegraphics[width=75mm]{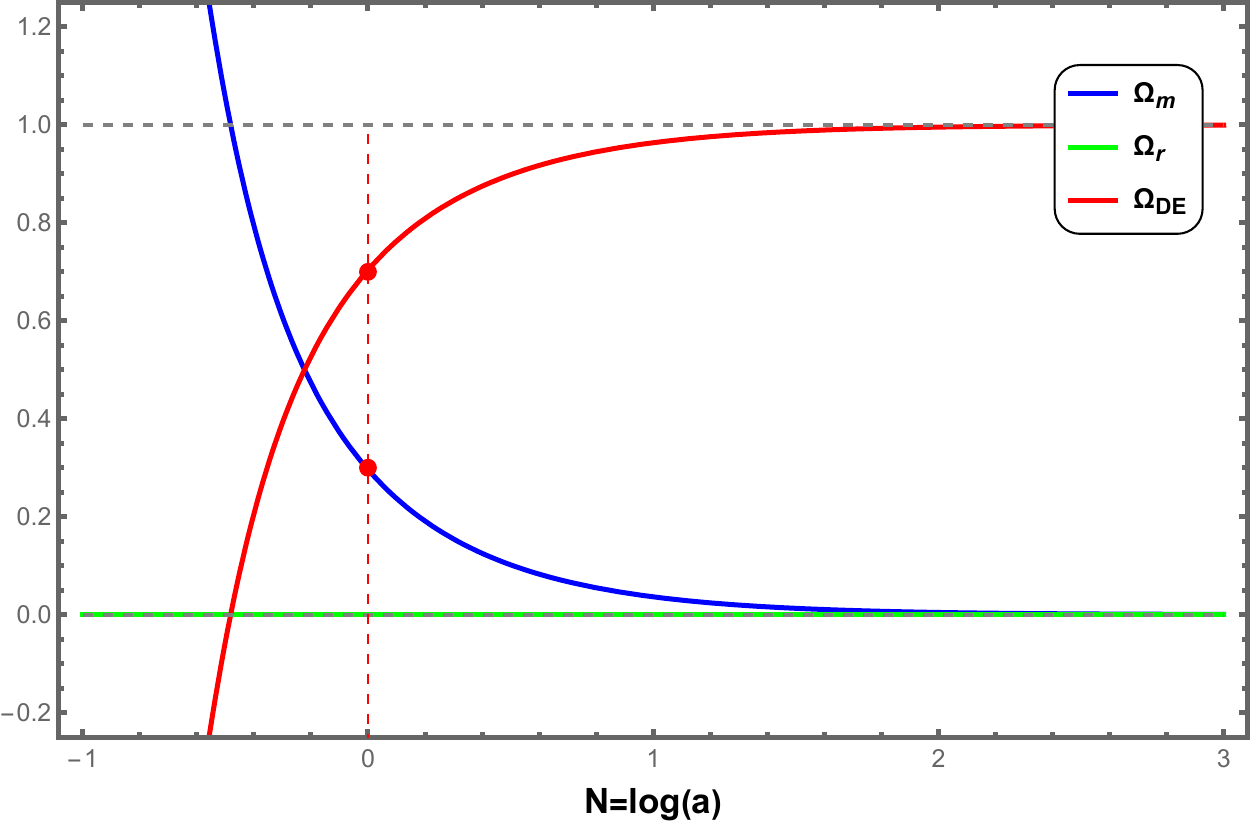}
    \caption{Evolution of the density parameters with initial conditions $X=-10^{-1},Y={10^{-5}},Z=10^{-10},V=10^{-11}$, $k=0.029$, $m=0.5$ (Model--I) } \label{fig:evolution I}
\end{figure}
\begin{figure}[H]
    \centering
    \includegraphics[width=75mm]{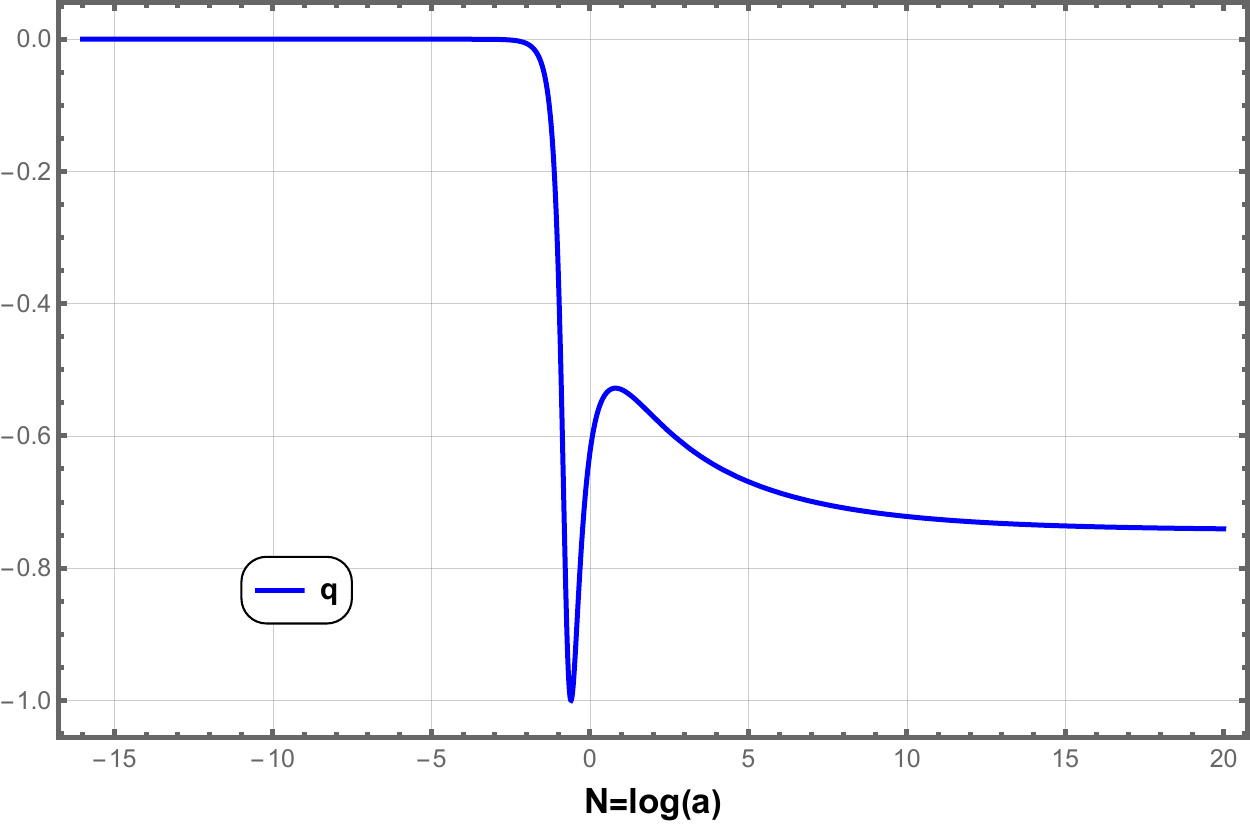}
    \includegraphics[width=75mm]{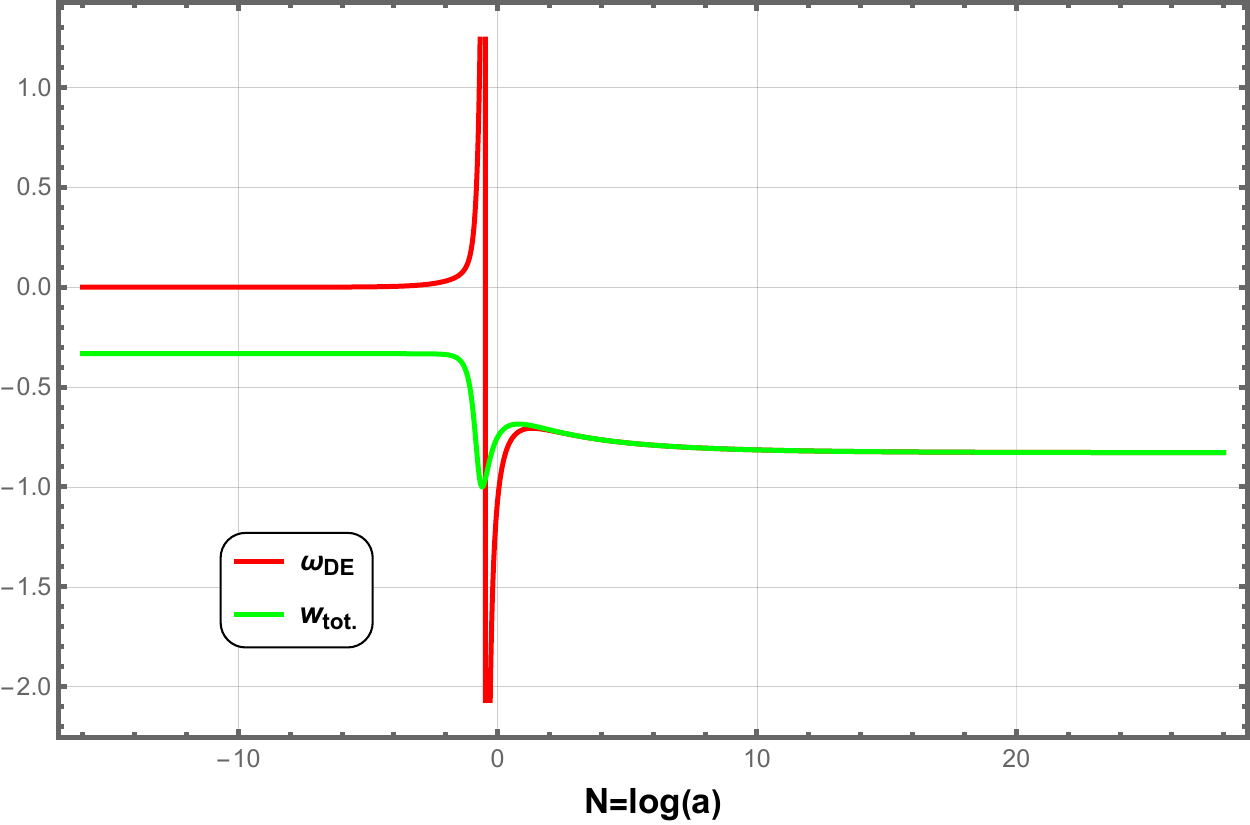}
    \caption{Deceleration and EoS parameter with initial conditions $X=-10^{-1},Y={10^{-5}},Z=10^{-10},V=10^{-11}$, $k=0.029$, $m=0.5$ (Model--I)} \label{Fig2}
\end{figure}

The evolution plot for the standard density parameter is plotted in Fig. \ref{fig:evolution I}. The vertical dashed line represents a present time of cosmic evolution, at which the standard density parameter for matter and DE shows values approximately equal to $0.3$ and $0.7$, respectively. The value of the deceleration parameter at present time is $q_{0}=-0.663$ which is compatible with the current observation study $q_0=-0.528^{+0.092}_{-0.088}$ \cite{Gruber:2014}. The plots for EoS parameters are presented in Fig. \ref{Fig2}, from which we can study the early and late time evolution of the Universe. The present value of $\omega_{DE}=-1.05$ which is compatible with the WMAP+CMB result [$\omega_0=-1.073 ^{+0.090}_{-0.089}$] \cite{Hinshaw:2013}.

%%%%%%%%%
\subsection{Sum of Separated Power Law Model}\label{Model-II}
We consider the sum of the separated power law \cite{Capozziello:2016eaz} form of $f(T, T_G)$ as
\begin{align}
f(T,T_G)=g_{0} T_{G}^{k} + t_{0}T^{m},
\label{secondtmodel}\end{align}
where $m$ is arbitrary and $k\neq 1$. In this case, from the dynamical system variables defined in Eq. (\ref{dynamical variables}), we find $f_{T}=-m W -\frac{4m}{k}\left(1+Z\right)X$. In this case, similar to Model--I, the dynamical variable $W$ can not be written as dependent on variables $X$ and $Z$; therefore, $W$ has to be treated as an independent variable, whereas the dynamical variable $Y$ shows the dependency on the variable $X$ as $Y=\frac{X \left(k-1\right) \left(2 Z^2 +\lambda +4Z \right)}{\left(Z+1\right)}$. In this case, the only term containing an explicit time-dependence form is the parameter $\lambda$, which plays an important role in identifying the particular epoch of the evolution of the Universe. The particular value of the parameter $\lambda$ can be reproduced by exact cosmological solutions representing a particular epoch of evolution. For $\lambda=0$, it can produce the exact de-Sitter scale factor, whereas, for $\lambda=\frac{9}{2}$, it is the matter-dominated exact solution. Further, the radiation-dominated scale factor reproduces for $\lambda=8$. The physical significance can be determined by the behavior of the EoS parameter ($\omega_{tot},\omega_{DE}$) and also the attracting solutions of this theory \cite{Odintsov:2018}. Following methods from the past literature, we consider the dynamical variable constant ($\lambda$) in the further analysis of this model \cite{Franco:2020lxx, Odintsov:2018, odintsov2017autonomous}. By referring to the general dynamical system defined in Eq. ($\ref{generaldynamicalsystem}$), we can write the dynamical system in autonomous form as,

\begin{align}
\frac{dX}{dN}&=\frac{(k-1) X (\lambda +2 Z (Z+2))}{Z+1}+2 X Z\,,\nonumber\\
\frac{dZ}{dN}&=\lambda-2Z^2\,,\nonumber\\
\frac{dV}{dN}&=-4V-2VZ \,,\nonumber\\
\frac{dW}{dN}&=\frac{8 m X Z (Z+1)}{k}+2 (m-1) W  Z-4 X (\lambda +2 Z (Z+2))\,.\label{DynamicalSystem2ndstmodel}
\end{align}

The density parameter for the DE and matter can be written as,

\begin{subequations}
\begin{align}
    \Omega_{DE}&=-2 m W +4 X Z+4 X+W-\frac{8 m X (Z+1)}{k}-\frac{4 (k-1) X (\lambda +2 Z (Z+2))}{Z+1}\,,\\
    \Omega_{m}&=1-V+2 m W -4 X Z-4 X-W +\frac{8 m X (Z+1)}{k}+\frac{4 (k-1) X (\lambda +2 Z (Z+2))}{Z+1}\,.
\end{align}
\end{subequations}
The critical points for the dynamical system in Eq. (\ref{DynamicalSystem2ndstmodel}) with the existing condition are presented in Table \ref{TABLE-V}.

\begin{table*} [!htb]
\small\addtolength{\tabcolsep}{-3pt}
\begin{tabular}{|*{7}{c|}}\hline
\parbox[c][1cm]{4cm}{\textbf{Name of Critical Points}} & \parbox[c][1cm]{0.5cm}{\textbf{X}} & \parbox[c][1cm]{0.5cm}{\textbf{Z}} & \parbox[c][1cm]{0.5cm}{\textbf{V}} & \parbox[c][1cm]{0.5cm}{\textbf{W}} & \parbox[c][1.5cm]{1.5cm}{\textbf{Exist for}} \\ 
\hline
\hline
\parbox[c][1cm]{4cm} {$B_1=(X_1,Z_1,V_1,W_1)$} & 0 & -2 & $V_1$ & 0 & $k\neq 0$, $\lambda=8$, arbitrary m, $V_1$. \\\hline
\parbox[c][1cm]{4cm} {$B_2=(X_2,Z_2,V_2,W_2)$} & $0$ & $-\frac{3}{2}$ & $0$ & $0$  & $\lambda=\frac{9}{2}, k \neq 0$. \\
\hline
\parbox[c][2cm]{4cm} {$B_3=(X_3,Z_3,V_3,W_3)$} & $X_3$ & $\epsilon_1$ & $0$ & $W_3$  & \begin{tabular}{@{}c@{}} $\epsilon_1 \ne 0, \epsilon_1+1\neq 0, k=\frac{1}{2},$\\ $W_3 =-8 (X_3 \epsilon_1+X_3), \lambda= 2 \epsilon_1^2$ \end{tabular}\\
\hline
\parbox[c][1cm]{4cm} {$B_4=(X_4,Z_4,V_4,W_4)$} & $0$ & $\epsilon_2$ & $0$ & $W_4$ & \begin{tabular}{@{}c@{}}$W_4  \epsilon_2\neq 0$,\\
$k \epsilon_2+k\neq 0, m=1$, $\lambda =2 \epsilon_2^2$.\end{tabular} \\
\hline
\end{tabular}
\caption{The critical points (Model--II)}
\label{TABLE-V}
\end{table*}

In this case, it can be noted that there are less number of critical points than in  Model--I; also, since $\lambda$ is an independent, we can categorize the critical points for different phases of evolution on the basis of the value of $\lambda$. The system will describe radiation, matter, and the DE-dominated era for $\lambda=8,\frac{9}{2}, \text{value depending on coordinates X and Y}$ respectively. The stability conditions, the values of the deceleration parameter, $\omega_{DE}, \omega_{tot}$ at each critical point are presented in Table \ref{TABLE-VI}.
\begin{table*}[!htb]
    % title of Table
    \centering % used for centering table
    \begin{tabular}{|*{5}{c|}}\hline
    \parbox[c][0.8cm]{1cm}{C. P.} & Stability Conditions & $q$ & $\omega_{tot}$ & $\omega_{DE}$ \\ [0.5ex] % inserts table %heading
    \hline\hline % inserts single horizontal line
    \parbox[c][0.6cm]{1cm}{$B_1$} & Unstable & $1$ & $\frac{1}{3}$ & - - \\
    \hline
    \parbox[c][0.6cm]{1cm}{$B_2$} & Unstable & $\frac{1}{2}$ & $0$ & - - \\
    \hline
    \parbox[c][0.6cm]{1cm}{$B_3$} &  \begin{tabular}{@{}c@{}} Stable for \\ $m<1\land \epsilon_1>0$ \end{tabular} & $-1-\epsilon_1$ & $-1-\frac{2 \epsilon_1}{3}$ & $\frac{2 \epsilon_1+3}{12 X_3-12 X_3 \epsilon_1}$ \\
    \hline
    \parbox[c][0.6cm]{1cm}{$B_4$} & \begin{tabular}{@{}c@{}}Stable for\\ $k<\frac{1}{2}\land \epsilon_2>0$ \end{tabular}& $-1-\epsilon_2$ & $-1-\frac{2 \epsilon_2}{3}$ & $\frac{-W_4 \epsilon_2+2 \epsilon_2+3}{3 W_4}$ \\
    \hline
    \end{tabular}
    \caption{Stability condition, deceleration and EoS parameter (Model--II)} 
    % is used to refer to this table in the text
    \label{TABLE-VI}
\end{table*}

The evolution equations and the standard density parameters for radiation, matter, and DE at each critical point are presented in Table \ref{TABLE-VII}.
\begin{table}
   % title of Table
    \centering % used for centering table
    \begin{tabular}{|*{6}{c|}}\hline
    \parbox[c][0.8cm]{1cm}{C. P.} & Evolution Eqs. & Universe phase & $\Omega_{r}$& $\Omega_{m}$ & $\Omega_{DE}$\\ [0.5ex] % inserts table %heading
    \hline\hline % inserts single horizontal line
    \parbox[c][0.7cm]{1cm}{$B_1$} & $\dot{H}=-2 H^{2}$ & $a(t)= t_{0} (2 t+c_{2})^\frac{1}{2}$ & $V_1$& $1-V_1$ & $0$\\
    \hline
    \parbox[c][0.7cm]{1cm}{$B_2$} & $\dot{H}=-\frac{3}{2}H^{2}$ & $a(t)= t_{0} (\frac{3}{2}t+c_{2})^\frac{2}{3}$ & $0$& $1$ & $0$\\
    \hline
    \parbox[c][0.7cm]{1cm}{$B_3$} & $\dot{H}=\epsilon_1 H^{2}$ & $a(t)= t_{0} (-\epsilon_1 t+c_{2})^\frac{-1}{\epsilon_1}$ & $0$& $1-4 X_3 \left(\epsilon_1-1\right)$ & $4 X_3 \left(\epsilon_1-1\right)$\\
    \hline
    \parbox[c][0.7cm]{1cm}{$B_4$} & $\dot{H}=\epsilon_2 H^{2}$ & $a(t)= t_{0} (-\epsilon_2 t+c_{2})^\frac{-1}{\epsilon_2}$ & $0$ & $W_4+1$ & $-W_4$\\
    
    \hline %inserts single line
    \end{tabular}
     \caption{Phase of the Universe, density parameters (Model--II)} 
     % is used to refer this table in the text
    \label{TABLE-VII}
\end{table}
The stability of the critical points is obtained from the signature of the eigenvalues and is presented in Table \ref{TABLE-VIII}. 

\begin{table}
\begin{tabular}{|*{2}{c|}}\hline
\parbox[c][0.5cm]{1cm}{C.P.} & \parbox[c][1cm]{8cm}{Eigenvalues}\\ \hline
\hline
\parbox[c][0.5cm]{1cm}{$B_1$} & \parbox[c][1cm]{8cm}{$\left\{0,8,-4 \left(m-1\right),-4 \left(2 k-1\right)\right\}$}\\ 
\hline
\parbox[c][0.5cm]{1cm}{$B_2$} & \parbox[c][1cm]{8cm}{$\left\{-1,6,-3 \left(2 k-1\right),-3 \left(m-1\right)\right\}$}\\
\hline
\parbox[c][0.5cm]{1cm}{$B_3$} & \parbox[c][1cm]{8cm}{$\left\{0,-4 \epsilon_1,2 \left(m-1\right) \epsilon_1,-2 \left(\epsilon_1+2\right)\right\}$}\\ 
\hline
\parbox[c][0.5cm]{1cm}{$B_4$} & \parbox[c][1cm]{8cm}{$\left\{0,-4 \epsilon_2,2 \left(2 k-1\right) \epsilon_2,-2 \left(\epsilon_2+2\right)\right\}$}\\ 
\hline
\end{tabular}
\caption{Eigenvalues corresponding to each critical point (Model--II)} % title of Table
\label{TABLE-VIII}
\end{table}

\begin{figure}[H]
    \centering
    \includegraphics[width=65mm]{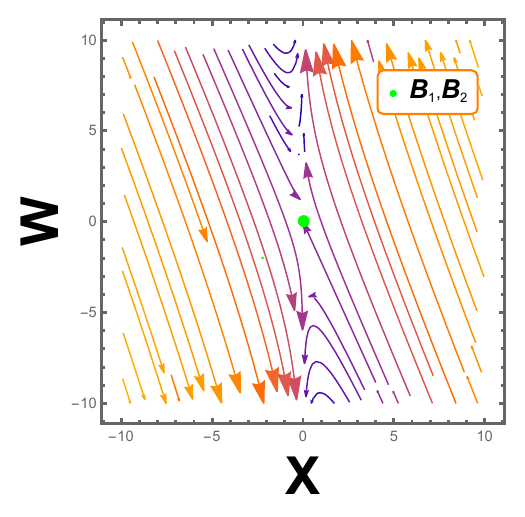}
    \includegraphics[width=65mm]{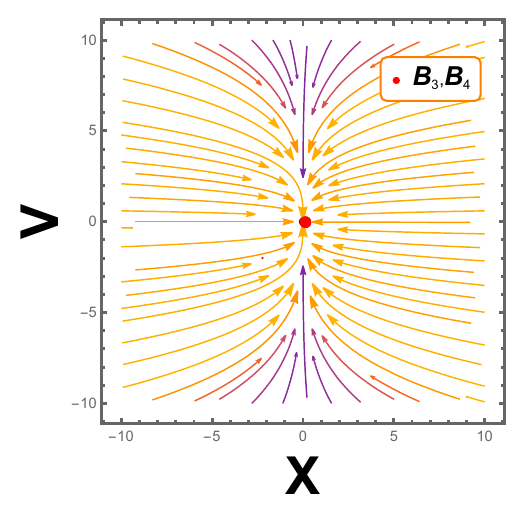}
    \caption{ $2D$ phase portrait for the dynamical system with $m=0.67$, $k=0.785$ ( Model--II) } \label{Fig3}
\end{figure}

The description for each critical point for the dynamical system in Eq. (\ref{DynamicalSystem2ndstmodel}) are as follow:

\textbf{Critical Point $B_1, (\lambda=8):$} The critical point $B_1$ is describing the radiation-dominated era with the parametric value of $Z=-2$, and $\lambda=8$. The value of deceleration parameter $q=1, \omega_{tot}=\frac{1}{3}$. The value of $\omega_{DE}$ is undetermined at $B_1$. From Table \ref{TABLE-VI}, one can observe that the critical points describing the radiation-dominated era are unstable. From Table \ref{TABLE-VIII}, in this case, the eigenvalues have zero as one of the eigenvalues; hence, these critical points are non-hyperbolic in nature. This critical point has at least one positive eigenvalue; hence, this critical point is unstable. The exact cosmological solution and the corresponding evolution equation are described in Table \ref{TABLE-VII}. The values of the standard density parameter imply that the critical point $B_1$ defines a standard radiation-dominated era at $V_1=1$. This critical point represents the non-standard radiation-dominated era in which a small amount of matter density parameter contributes. The phase space trajectories near this critical point can be analyzed in Fig. \ref{Fig3}. The critical point representing the radiation-dominated era is plotted in a single plot. The phase space trajectories are moving away from all these critical points; hence, we can analyze the saddle point behavior, which is unstable.

\textbf{Critical Points $B_2, (Z=-\frac{3}{2}, \lambda=\frac{9}{2})$:}  This critical point is describing the cold-dark matter-dominated era. From Table \ref{TABLE-VII}, we can observe that the critical point $B_{2}$ represents the standard cold dark matter-dominated era. Again from Table \ref{TABLE-VI}, we see the value of $q=\frac{1}{2}$ and $\omega_{tot}=0$ at this critical point. From Table \ref{TABLE-VIII}, we observe that the eigenvalues are unstable, and the existence of one zero eigenvalue at this critical point implies that this critical point is non-hyperbolic. The behavior of phase space trajectories can be analyzed from Fig. \ref{Fig3}. The trajectories at these critical points move away from the critical points; hence, critical points show saddle point behavior.

\textbf{Critical Points $B_3 \text{-} B_{4}$, ($\lambda,~Z $ depend on free parameters $\epsilon_1$ and $\epsilon_2$):} In these critical points, the value of parameter $\lambda$ and the parameter $Z$ is depend on $\epsilon_1$ and $\epsilon_2$. The critical point  $B_3$ and at $B_{4}$ may describe current cosmic acceleration, respectively, at $\epsilon_1>-1$ and $\epsilon_2>-1$. The de-Sitter solution can be analyzed at $\epsilon_1, \epsilon_2=0$ respectively for  critical points $B_{3}$ and $B_{4}$. The hyperbolic or non-hyperbolic nature of these critical points depends on the coordinate values and shows stability at the conditions described in Table \ref{TABLE-VI}. The exact cosmological solution and the evolution equations at these critical points are described in Table \ref{TABLE-VII}. From the behavior of phase space trajectories in Fig. \ref{Fig3}, it can be observed that phase space trajectories are attracting towards both the critical points; hence, these critical points are attractors in behavior.
\begin{figure}[H]
    \centering
    \includegraphics[width=70mm]{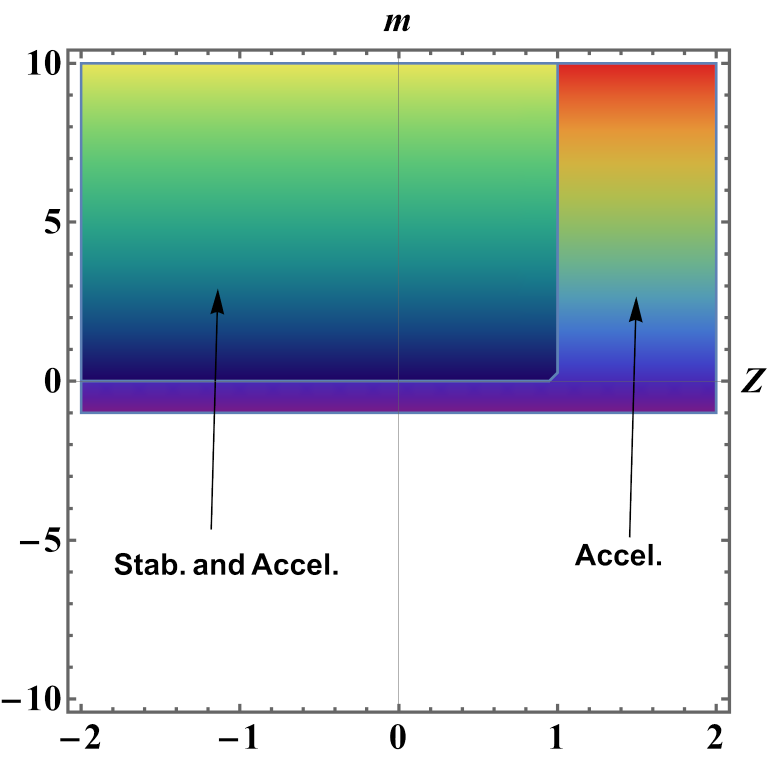}
    \includegraphics[width=70mm]{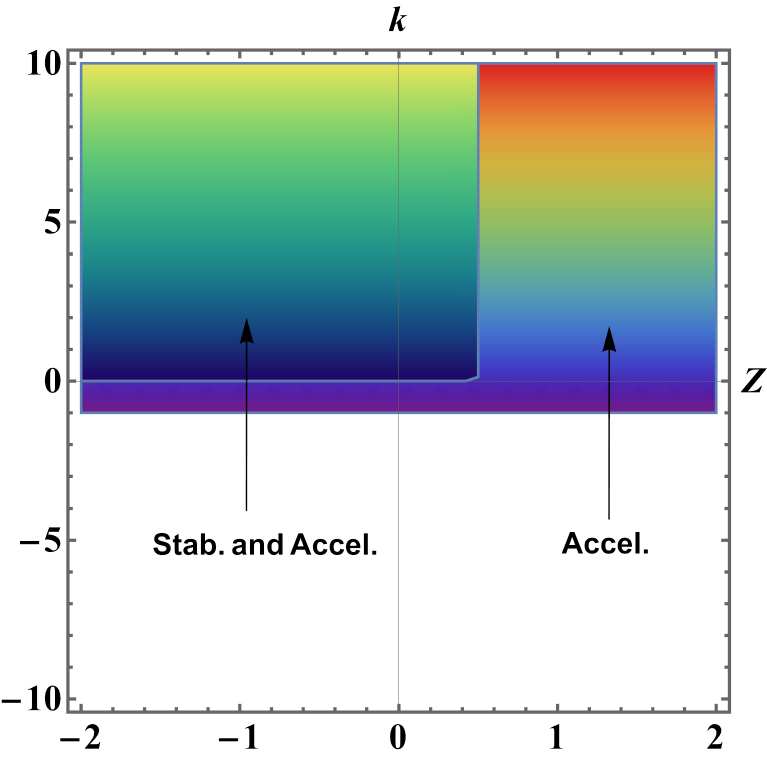}
    \caption{Region Plots for critical points $B_3$ and $B_4$ showing stability and acceleration along with ranges of the model parameters $m,k$ vs dynamical variable $Z$. } \label{regionplotm2}
\end{figure}
We wish to mention here that the fixed points $B_1$ and $B_2$ exist only for the value of $\lambda$ as $8$ and $\frac{9}{2}$, respectively. For the choice of the variables of the model, all three different epochs of the evolution of the Universe could not be obtained in a single phase space for fixed $\lambda$. Fig. \ref{regionplotm2} describes the critical points $B_3$ and $B_4$ are stable and show accelerating behavior for the range of the model parameters $\left((m < 1) \wedge (z > 0)\right) \wedge \left((k<\frac{1}{2}) \wedge (z > 0)\right)$.\\

\begin{figure}[H]
    \centering
    \includegraphics[width=75mm]{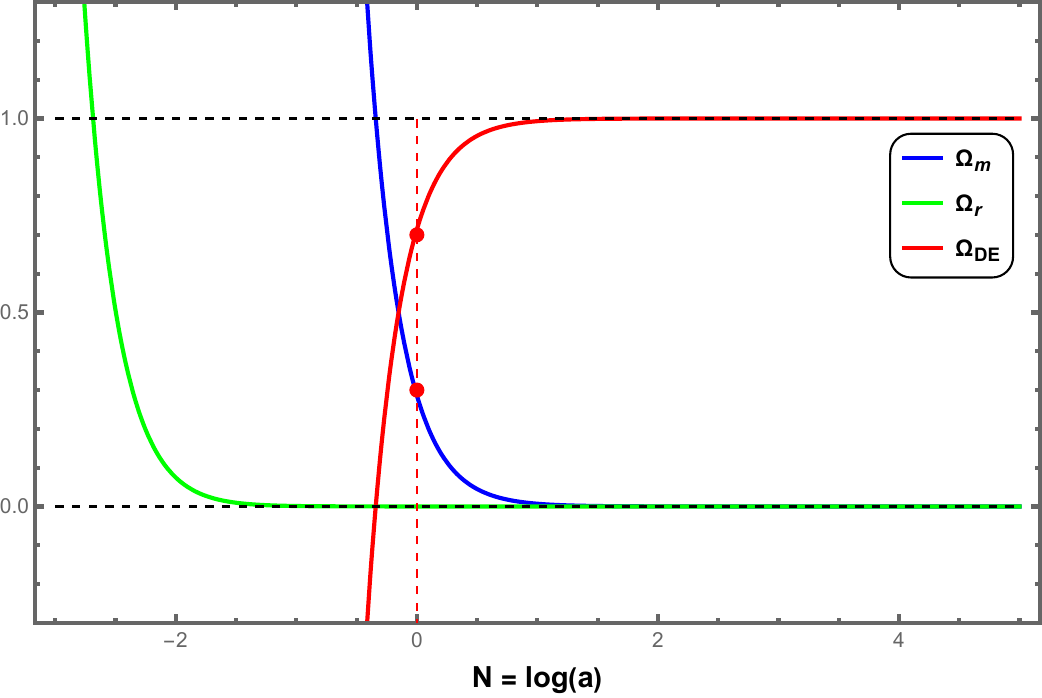}
    \caption{Evolution of the density parameters with initial conditions $X=10^{-1.3},\, Z=0.02 \times 10^{1.2},\, V=0.02 \times 10^{-2.5},\, W=0.0021 \times 10^{-3.7}$,\, $ m=0.67$,\, $k=0.785$ ( Model--II)} \label{fig:evolution II}
\end{figure}
\begin{figure}[H]
    \centering
    \includegraphics[width=75mm]{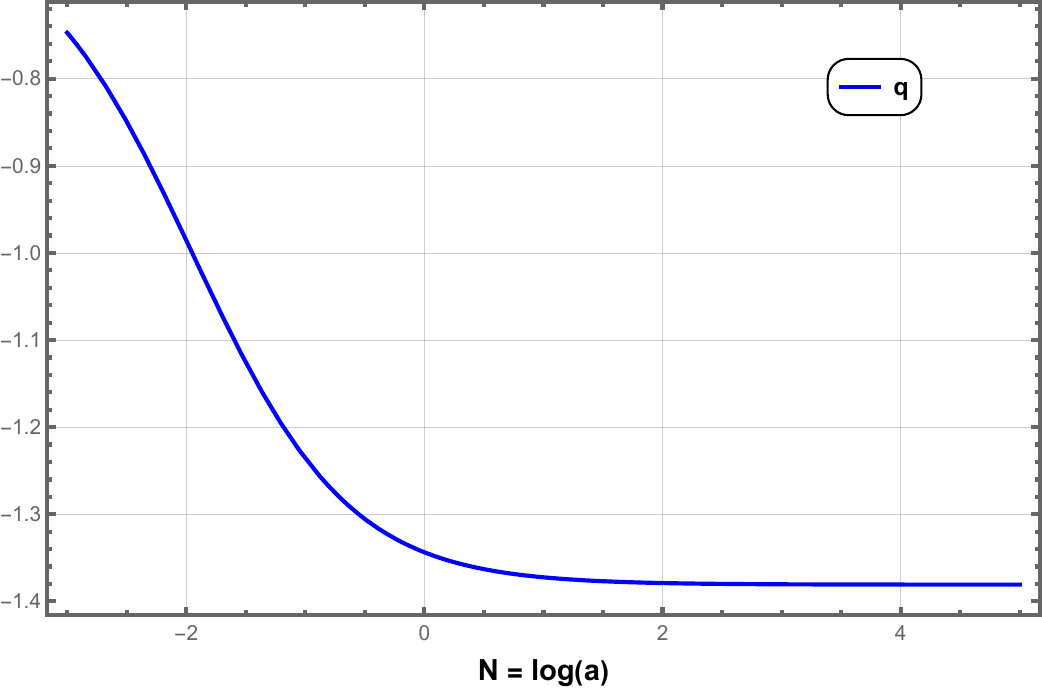}
    \includegraphics[width=75mm]{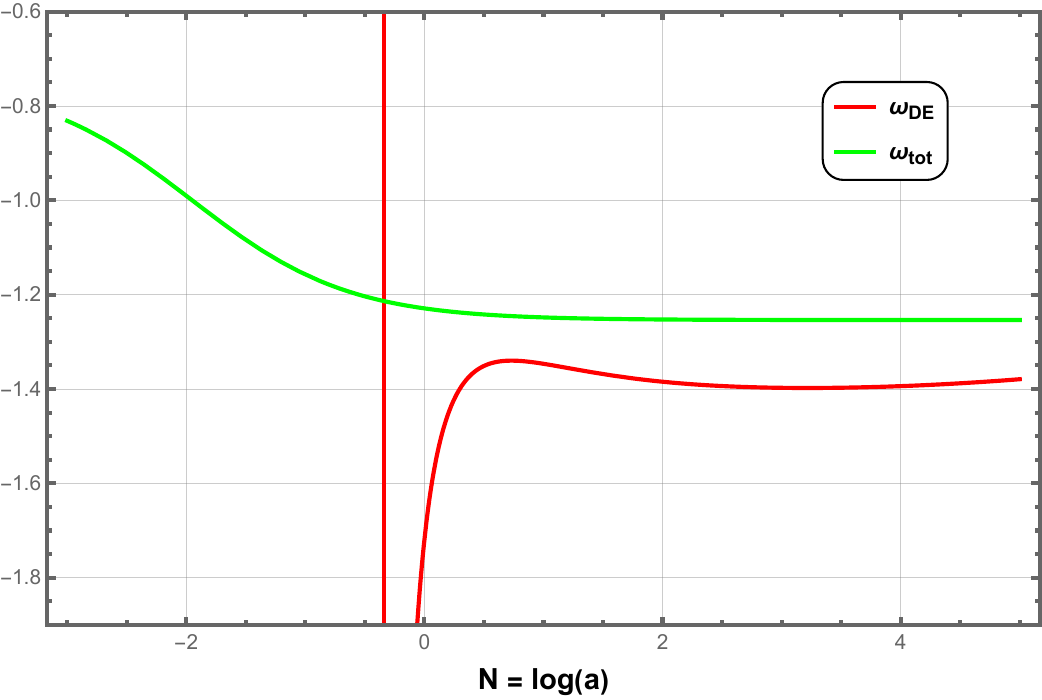}
    \caption{Deceleration and EoS parameter with initial conditions $X=10^{-1.3},\,  Z=0.02 \times 10^{1.2},\, V=0.02 \times 10^{-2.5},\, W=0.0021 \times 10^{-3.7}$,\, $ m=0.67$,\, $k=0.785$ (Model--II).} \label{Fig4}
\end{figure}

The evolution of density parameters $\Omega_{r}$, $\Omega_{m}$, $\Omega_{DE}$ have been presented in Fig. \ref{fig:evolution II}. The vertical dashed line represents the present time at which the values of $\Omega_{DE} \approx 0.7$ and $\Omega_{m} \approx 0.3.$ At the early epoch, we can see that the evolution curve for $\Omega_{r}$ is dominating the other two curves, but it will go on decreasing from the early to the late time of cosmic evolution. The deceleration parameter $q$ and EoS parameters in redshift $N = log(a)$ have been given in Fig. \ref{Fig4}. Currently, the value of the deceleration parameter is obtained as $q_{0}=-1.345$, which agrees with the range provided in Ref. \cite{Feeney:2018}. The present value of the DE EoS parameter has been obtained as $\omega_{DE}=-1.08$ and is approximately the same as in Ref. \cite{Hinshaw:2013}.

\section{Summary and Conclusion}\label{sec:conclusion}

 In this paper, we have performed the dynamical system analysis in a modified $f(T, T_G)$ gravity framework. Two well-motivated forms of $f(T, T_G)$ are considered, such as (i) mixed power law and (ii) sum of the separated power law. It is possible to select functions which assign the models according to the Noether symmetry approach, and then those functions (and then the models) should be physically applicable. The critical points of each model have been obtained, as well as its stability behavior has been examined. In the first model, the critical points, $A_1,\, A_2,\, A_3,\, A_4,$ and $A_5$, are obtained for some specific ranges of the model parameters. The critical points $A_3$, $A_4$, and $A_5$ are showing stable behavior. The range of model parameters at which the stability and accelerating behavior of $A_3$ and $A_4$ can be achieved as $\left((\frac{1}{4} < k \leq \frac{13}{25}) \wedge \left(m > \frac{1}{6}\right)\right) \vee \left((\frac{1}{4} < k \leq \frac{13}{25}) \wedge \left(m \leq -\frac{11}{6}\right)\right)$. Whereas the stability of critical point $A_5$ is dynamical variable dependent, it has a significant role in analyzing the early and future fate of the Universe. From the evolution plot, the present value of the density parameter obtained to be $\Omega_{DE} \approx 0.7$ and $\Omega_{m} \approx 0.3$. The present value of deceleration and DE EoS parameters has been obtained respectively as $q=-0.663$ and $\omega_{DE}=-1.05$. In the second model, we have obtained four critical points such as    $B_1,\, B_2,\, B_3,\,$ and $B_4$. The two stable points $B_3$ and $B_4$ are showing accelerating and stable behavior in the range of the model parameters  $\left((m < 1) \wedge (z > 0)\right) \wedge \left((k<\frac{1}{2}) \wedge (z>0)\right)$. For this model, the present value of density parameters is obtained as $\Omega_{DE} \approx 0.7$ and $\Omega_{m} \approx 0.3$. Further, the present value of deceleration and DE EoS parameters are obtained to be $q=-1.345$ and $\omega_{DE}=-1.08$.    

The dynamical variable $\lambda$ has been crucial in identifying the phases of the evolution of the Universe. In the first model, the phase space trajectories pass through $A_1 \rightarrow A_2 \rightarrow A_4, A_5$, representing the evolution from radiation to matter to DE era. In the second model for a fixed value of $\lambda$, the evolution phases could not be established in a single-phase space.  Our study highlights the significance of dynamical system analysis in $f(T, T_G)$ gravity, showcasing the ability to obtain the present value of the deceleration parameter, EoS parameter, and density parameters, which are compatible with the current cosmological observations. To distinguish this study from $f(R, G)$ gravity, we can see that a greater number of critical points can be obtained in $f(T, T_G)$ gravity to explain the DE phase. The function $f(T, T_G)$ has a rather complex form as it contains a square root in \cite{Kofinas:2014daa}, which studies the cosmological implications of $f(T, T_G)$ gravity as a dynamical system. In order to provide novel mass scales, we have examined the dynamical system for the mixed power-law and the sum of the separated power-law forms of the functions. The Noether symmetry approach allows us to fix the form of the function $f(T, T_G)$ and to derive exact cosmological solutions \cite{Capozziello:2016eaz}. However, even though this model appears to have an intact dynamical system, it has not yet been thoroughly tested against observations such as Supernova, CMB, BAO, and so on, and the local gravity constraints have to be addressed and shown to be under control in order to make it an effective model.

\section*{Acknowledgements}
SAK acknowledges the financial support provided by the University Grants Commission (UGC) through Senior Research Fellowship (UGC Ref. No.: 191620205335), and S.V.L. acknowledges the financial support provided by the University Grants Commission (UGC) through Senior Research Fellowship (UGC Ref. No.: 191620116597) to carry out the research work. BM acknowledges IUCAA, Pune, India, for hospitality and support during an academic visit where a part of this work has been accomplished. 
\section*{References} 
\bibliographystyle{utphys}
\bibliography{references}

\end{document}